\documentclass[10pt,journal,compsoc]{IEEEtran}
\pdfoutput=1
\usepackage{makecell}
\usepackage{graphicx}
\usepackage{framed,subfigure}
\usepackage[ruled,linesnumbered,vlined]{algorithm2e}
\newtheorem{myDef}{Definition}
\newtheorem{myEg}{Example}
\begin{document}

\title{A Dual-Store Structure for Knowledge Graphs}

\author{Zhixin Qi,
        Hongzhi Wang,~\IEEEmembership{Member,~IEEE,}
        and~Haoran Zhang
\IEEEcompsocitemizethanks{\IEEEcompsocthanksitem Z. Qi is with the School of Computer Science and Technology, Harbin Institute of Technology, P.R. China.
\IEEEcompsocthanksitem H. Wang is with the School of Computer Science and Technology, Harbin Institute of Technology, P.R. China. Email: wangzh@hit.edu.cn. Corresponding author.
\IEEEcompsocthanksitem H. Zhang is with the School of Software, Harbin Institute of Technology, P.R. China. }
\thanks{Manuscript received April 19, 2020; revised August 26, 2020.}}

\markboth{Journal of \LaTeX\ Class Files,~Vol.~14, No.~8, August~2020}%
{Shell \MakeLowercase{\textit{et al.}}: Bare Demo of IEEEtran.cls for Computer Society Journals}

\IEEEtitleabstractindextext{
\begin{abstract}
To effectively manage increasing knowledge graphs in various domains, a hot research topic, knowledge graph storage management, has emerged. Existing methods are classified to relational stores and native graph stores. Relational stores are able to store large-scale knowledge graphs and convenient in updating knowledge, but the query performance weakens obviously when the selectivity of a knowledge graph query is large. Native graph stores are efficient in processing complex knowledge graph queries due to its index-free adjacent property, but they are inapplicable to manage a large-scale knowledge graph due to limited storage budgets or inflexible updating process. Motivated by this, we propose a dual-store structure which leverages a graph store to accelerate the complex query process in the relational store. However, it is challenging to determine \emph{what} data to transfer from relational store to graph store at \emph{what} time. To address this problem, we formulate it as a Markov Decision Process and derive a physical design tuner \textsf{DOTIL} based on reinforcement learning. With \textsf{DOTIL}, the dual-store structure is adaptive to dynamic changing workloads. Experimental results on real knowledge graphs demonstrate that our proposed dual-store structure improves query performance up to average 43.72\% compared with the most commonly used relational stores.
\end{abstract}

}

\maketitle

\IEEEdisplaynontitleabstractindextext

\IEEEpeerreviewmaketitle

\section{Introduction}
\label{sec:introduction}
\IEEEPARstart{A}{s} the cornerstone of artificial intelligence, knowledge graphs play an essential role in enhancing the cognitive ability of machines. A great number of knowledge graphs have been constructed in various domains, such as DBpedia~\cite{dbpedia} integrated by the resources of Wikipedia, geographical knowledge base LinkedGeoData~\cite{linkedgeodata}, and UniProt~\cite{uniprot} collected by the information on proteins. To take advantage of knowledge graphs, effective storage management is identified as the basic premise. In real applications, it is challenging to store knowledge graphs effectively due to the following characteristics of a knowledge graph.

\textbf{$\bullet$ Large scale.} A knowledge graph often contains millions of vertices and billions of edges. For instance, the number of triples in DBpedia is greater than 3 billion, there are about 3 billion edges in LinkedGeoData, and UniProt has more than 13 billion triples. The large scale makes it difficult to store the entire knowledge graph in a centralized graph database. For example, it is intractable to manage an entire knowledge graph in Neo4j database due to its cumbersome importing process~\cite{neo4j}. Even if there are few data changes, users have to reload the entire graph and restart the database. Also, since the storage constraint of a standalone gStore database is 1 billion triples, it is impossible to store all the triples of a large-scale knowledge graph~\cite{zou2014gstore}.

\textbf{$\bullet$ Complex Query Patterns.} Since the nodes and edges of a knowledge graph represent entities and relationships in the real world, a knowledge graph contains various semantics in different domains. To make full use of the semantic information, the patterns of knowledge graph queries are usually complex. In this paper, complex query patterns refer to the query patterns containing more than one predicate. Such as ``students who took the same course'', ``actors who acted in the same movie'', and ``people who was born in the same city as his or her advisor''. These complex queries bring a heavy burden on relational databases for graph data. To answer them, the relational tables perform multiple join operations and produce expensive time costs, while the index-free adjacent property of graph databases makes the time complexity of graph traversal positively related to the traversal range but irrelevant to the entire graph size~\cite{Rodriguez2012The}.

\textbf{$\bullet$ Changing Data and Workloads.} The frequency of updating a knowledge graph and its workload becomes higher as the number of users increases. This property brings two challenges to knowledge graph stores. On the one hand, the increasing knowledge requires high efficiency in inserting new data into the database. On the other hand, in order to guarantee the efficiency of query process, the storage structure needs to adjust itself according to the dynamic workloads.

These properties of knowledge graph data have great impacts on the storage effectiveness and bring the above challenges to the knowledge graph store design. Unfortunately, existing stores are unable to tackle them. Current storage structures for knowledge graphs are classified into two lines, i.e. relation-based stores~\cite{3store,DLDB,Jena,abadi2009sw,neumann2008rdf,weiss2008hexastore,sun2015sqlgraph} and native graph stores~\cite{neo4j,zou2014gstore}. Even though they are efficient for some knowledge graphs, their effectiveness is not satisfactory in many scenarios. Then, we give detailed discussions.

Relation-based stores are able to store large-scale knowledge graphs and insert new triples efficiently. However, when the selectivity of a complex knowledge graph query is large, relational databases answer the query by scanning the tables instead of using indexes, which hurts the query performance severely. From the aspect of native graph stores, they are efficient to answer complex queries by index-free adjacent property~\cite{Rodriguez2012The}. However, they take a long time to insert new edges or vertexes and usually have a storage constraint~\cite{pokorny2017integrity}. In addition, both of these two stores are static. They cannot adjust themselves automatically with the changing patterns of their workloads.

Motivated by these, we attempt to design a novel knowledge graph storage structure which overcomes the drawbacks of existing stores. We compare the performances of a representative relational database MySQL and a popular graph database Neo4j by answering a complex SPARQL query ``SELECT ?p WHERE\{?p y:wasBornIn ?city. ?p y:hasAcademicAdvisor ?a. ?a y:wasBornIn ?city.\}''. To test the relationship between query time and data size, we vary the number of knowledge graph triples in the two databases from 500,000 to 5,000,000. As Table~\ref{table:pre-exp} shows, as the data size rises, the query time of MySQL becomes longer significantly while the time cost of Neo4j grows slowly. We observe that the larger data size of a knowledge graph has little impact on the complex query performance of Neo4j. Since Neo4j is efficient to answer complex queries but has a storage constraint~\cite{pokorny2017integrity}, while MySQL is able to store large-scale knowledge graphs, the goal of our physical design is to leverage the specific benefits of the two stores.

\begin{table}[!htb]
	\centering
	\caption{Query Performances of MySQL and Neo4j Varying \#-triples from 500000 to 5000000 (Unit: s)}
	\label{table:pre-exp}
	\begin{tabular}{cccccc}
		\Xhline{0.8pt}
		 & 500000 & 1000000 & 1500000 & 2000000 & 2500000  \\
		\hline
		MySQL & 11.2304 & 17.2368 & 27.6332 & 37.6454 &  47.9656  \\
		Neo4j & 0.6067 &  1.3270 & 1.5837 & 3.3893 &  2.2573   \\
	   \Xhline{0.8pt}
		& 3000000 & 3500000 & 4000000 & 4500000 & 5000000  \\
		\hline
		MySQL & 62.5006 & 69.7482 & 68.8358 & 68.6312 & 99.4103   \\
		Neo4j & 3.4786 & 2.7923 & 3.4560  & 3.7312 &  3.9833   \\
		\Xhline{0.8pt}
	\end{tabular}
\end{table}

To achieve this goal, we explore a dual-store structure for knowledge graphs, i.e. using the graph database as an accelerator for knowledge graph complex queries, while preserving the above-mentioned roles of both stores. The dual-store structure combines the advantages of relation-based and native graph stores by enabling the relational database to store the entire data and utilize the graph database for complex query processing. Naturally, this structure requires the graph database to store the share of data for complex queries answering. However, it is challenging in that the knowledge graph workload is changing over time. The crucial problem for it is to determine \emph{what} data to transfer from the relational database to the graph database at \emph{what} time. We refer to this problem as the physical design tuning of a dual-store structure.

While physical design tuning is a well-known problem, prior researches consider either a single store or a multi-store including relational store, key-value store, document store, HDFS, and NoSQL~\cite{bruno2007online,schnaitter2007line,schnaitter2012semi,consens2012divergent,lefevre2014miso,bugiotti2015invisible}. The existing techniques could not be applied to the physical design tuning problem of our dual-store structure. On the one hand, frequent changes of knowledge graph and its workload require the dual-store tuner to tune the physical design automatically based on historical experience. However, existing heuristic-based methods lack the ability to memorize effectiveness of different tuning operations, and hence cannot accumulate historical experiences to guide future physical design tuning. On the other hand, the majority of current studies treat materialized views or indexes as the basic elements of physical design. However, in our problem, the design elements are data partitions in knowledge graphs. Thus, to the best our knowledge, this is the first work to tune the dual store with relational database and native graph database to cope with large graph management.

We treat this problem as a variant of knapsack problem~\cite{horowitz1974computing}. Since the benefit of each data partition is unknown and changing, this problem is more difficult than the knapsack problem. Thus, the dual-store physical design tuning problem is at least NP-hard. To solve our problem, we need to memorize the previous benefits of data partitions from the historical experiences. Inspired by the memory ability of machine learning, we tune the dual-store structure with the help of machine learning models.

Since each tuning operation is affected by the results of last operation, we model the solution of our problem as a Markov Decision Process (MDP for brief). Given the current state of the dual-store structure, the goal is to decide the next action. That is, which data partitions are valuable to transfer from the relational database to the graph database. Naturally, we adopt reinforcement learning (RL for brief) techniques to optimize this MDP objective based on the observed performance and previous experience~\cite{van2017automatic,kara2018columnml,kraska2018case,ding2019ai}. Assuming there are $n$ data partitions in a knowledge graph, the state space is $2^n$. The difficulty of RL-based solution is how to train the RL model effectively with a large state space and limited workloads. To tackle this difficulty, we incorporates a state space decomposition strategy and a counterfactual scenario into our proposed RL-based dual-store tuner \emph{DOTIL}. The decomposition strategy decomposes the state space and increases the retraining frequency of each state, and the counterfactual scenario guarantees the utilization of each complex query in the workloads.

In summary, the main contributions of this paper are listed as follows.

$\bullet$ We provide a dual-store structure for knowledge graphs, which has the unique combination of features: $i$) ability to store large-scale knowledge graphs; $ii$) high-efficiency in complex query processing and data insertion; $iii$) adaptivity to the changing workloads.

$\bullet$ For the efficiency and adaptivity of the dual-store structure, we study the physical design tuning problem. To the best of our knowledge, this is the first work to tune the physical design which consists of relation-based and native graph stores.

$\bullet$ We propose \emph{DOTIL}, a Dual-stOre Tuner based on reInforcement Learning, to automatically determine which data partitions to transfer from the relation-based store to the graph store at what time according to the dynamic workloads. DOTIL not only guarantees the adaptivity of our dual-store structure, but also soups up the complex query processing.

$\bullet$ To demonstrate the effectiveness of our store, we conduct extensive evaluations on real knowledge graphs. Evaluation results show that the dual-store structure improves the query performance up to average 43.72\% compared with the most commonly used relational store, and up to average 63.01\% compared with the relational store optimized by materialized views.

The rest of the paper is organized as follows. We review related work in Section~\ref{sec:related}. Section~\ref{sec:frame} introduces our proposed dual-store structure. We discuss the physical design tuning problem and our reinforcement-learning-based solution in Section~\ref{sec:dotil}. Section~\ref{sec:plan} explains the query processor in dual-store structure. Experiments are conducted in Section~\ref{sec:exp}. Section~\ref{sec:conclude} concludes this paper.

\section{Related Work}
\label{sec:related}
The related work is divided into the following main classes.

\textbf{Knowledge Graph Stores.} Many storage structures have been proposed for knowledge graphs and are classified into two lines, that are relation-based stores~\cite{3store,DLDB,Jena,abadi2009sw,neumann2008rdf,weiss2008hexastore,sun2015sqlgraph} and native graph stores~\cite{neo4j,zou2014gstore}.

Triple table is the earliest knowledge graph store~\cite{3store}. Each tuple of it stores the subject, predicate, and object of an edge in knowledge graphs. This store is convenient for users, but produces a lot of self-join operations during the query process. Horizontal table stores all the predicates and objects of a subject in a tuple~\cite{DLDB}. The column number of it is the amount of predicates in knowledge graphs, while the row number is the amount of subjects. It generates less self-join operators than triple table store, but it has more disadvantages. First, since there are thousands of predicates in a large-scale knowledge graph, the amount of predicates may exceed the limit of column number in a relational database. Second, different subjects own different predicates. Thus, there are a lot of null values in the database. Third, it is costly to update the knowledge graph in that the table structure changes when inserting, modifying, or deleting a column. To reduce the great column number of horizontal table, property table stores the subjects belonging to the same type in a table~\cite{Jena}. Since it produces multi-table joins rather than self-table joins, the query readability is improved. However, the table number may exceed the constraint of relational database due to the thousands of subjects in a knowledge graph. Moreover, there are still many null values since the predicate set of different subjects are different. Vertical partitioning creates a two-column table storing the related subjects and objects for each predicate~\cite{abadi2009sw}. This store eliminates the null values in the tables, but produces a great many of multi-table joins when processing complex queries, especially the queries without given predicates. Sextuple indexing is an extension of triple table~\cite{neumann2008rdf,weiss2008hexastore}. It creates six tables for all the permutations of triples in a knowledge graph. This store ameliorates the brittleness of self-joins and speeds up the query process, but still has two drawbacks. One is the high costs of storage spaces. As the scale of a knowledge graph becomes larger, the storage expenses rise quickly. The other is the large number of join operations between index tables in the case of complex knowledge graph queries. To trade-off the characteristics of triple table, property table, and vertical partitioning, DB2RDF creates dph (direct primary hash) table, ds (direct secondary hash) table, rph (reverse primary hash) table, and rs (reverse secondary hash) table~\cite{sun2015sqlgraph}. The dph table stores a subject, its predicates and the corresponding objects in each tuple, the ds table records the object values of multi-value predicates in the dph table, the rph table stores an object, its predicates, and the corresponding subjects in each tuple, and the rs table contains the subject values of multi-value predicates in the rph table. Although this store accelerates the knowledge graph query process by reducing the number of join operations and null values, it lacks self-adaptivity to the changing workloads.

The native graph stores for knowledge graphs are less mature than the relation-based stores. Neo4j is the most popular property graph database~\cite{neo4j}. Due to its index-free adjacency, the query performance on it is irrelevant to the graph size, and only related to the range of graph traversal. In spite of the high-efficiency in complex query processing, Neo4j is costly to store an entire knowledge graph due to its cumbersome importing process. When there are changes in the knowledge graph, users need to reload the entire graph and restart Neo4j. Another representative graph database gStore adopts a bit-string-based storage strategy~\cite{zou2014gstore}. It maps the attributes and attribute values of each resource in an RDF graph to a binary bit string, and creates a VS* tree with the bit strings. During the execution of SPARQL queries, the VS* tree is used as an index to find the query variables. Even though the index speeds up the query process, gStore is unable to store a large-scale knowledge graph due to its storage limit. 

Compared with existing knowledge graph stores, our proposed dual-store structure has two advantages: $i$) Instead of storing knowledge graphs in a single store, we explore a dual-store structure which combines the strengths of both relational-based and native graph stores. Our store is not only able to store a large-scale knowledge graph, but also high-efficiency in complex query processing and knowledge graph updating. $ii$) Since the query workloads are changing over time, the adaptivity to dynamic workloads is an essential property of a knowledge graph store. However, none of existing literatures considers this property. To fill this gap, our dual-store structure adopts a reinforcement-learning-based tuner to adjust the physical design of our store according to the changing workloads.

\textbf{Physical Design Tuning.} Majority of the early works focus on physical design tuning for centralized RDBMS systems~\cite{bruno2007online,schnaitter2007line,schnaitter2012semi}. The objective is to minimize the execution cost of a dynamically changing workload. New designs are materialized in parallel with query execution or during a tuning phase. All these works consider a single data store and use indexes as the elements of physical design. Later, Consens et al. study the tuning problem for replicated databases, where each RDBMS is identical and the workload is provided offline~\cite{consens2012divergent}. LeFevre et al. tune a multistore system online and utilize opportunistic materialized views as the design elements~\cite{lefevre2014miso}. The multistore system involves two types of data stores, that are a parallel relational data warehouse and a system for massive data storage and analysis (namely HDFS with Apache Hive). Bugiotti et al. propose a self-tuning multistore architecture including NoSQL system, key-value store, document store, nested relations store, and relational store. The architecture utilizes view-based rewriting and view selection algorithms to correctly handle the features of diverse data models involved~\cite{bugiotti2015invisible}.  

Comparing our tuner DOTIL with existing studies of physical design tuning, we find that: $i$) Since none of prior work studies the physical design tuning of relational and graph stores, our paper aims to address this problem. $ii$) Instead of relying on heuristics, DOTIL uses the ``working by doing'' mechanism of reinforcement learning to tune the dual-store structure for knowledge graphs. By updating its model continuously and automatically based on delayed observations and historical experience, DOTIL makes our store adaptive to the dynamic workloads.

\textbf{Summary.} Our study differentiates itself from all these studies in two aspects. On the one hand, none of existing literatures considers a dual-store structure for knowledge graphs. On the other hand, the study of physical design tuning of relation-based and native graph stores is unexplored in previous work.
\section{Overview}
\label{sec:frame}
In this section, we introduce the overview of our proposed dual-store structure.

\begin{figure}
	\centering
	\includegraphics[width=2.3in,height=2.4in]{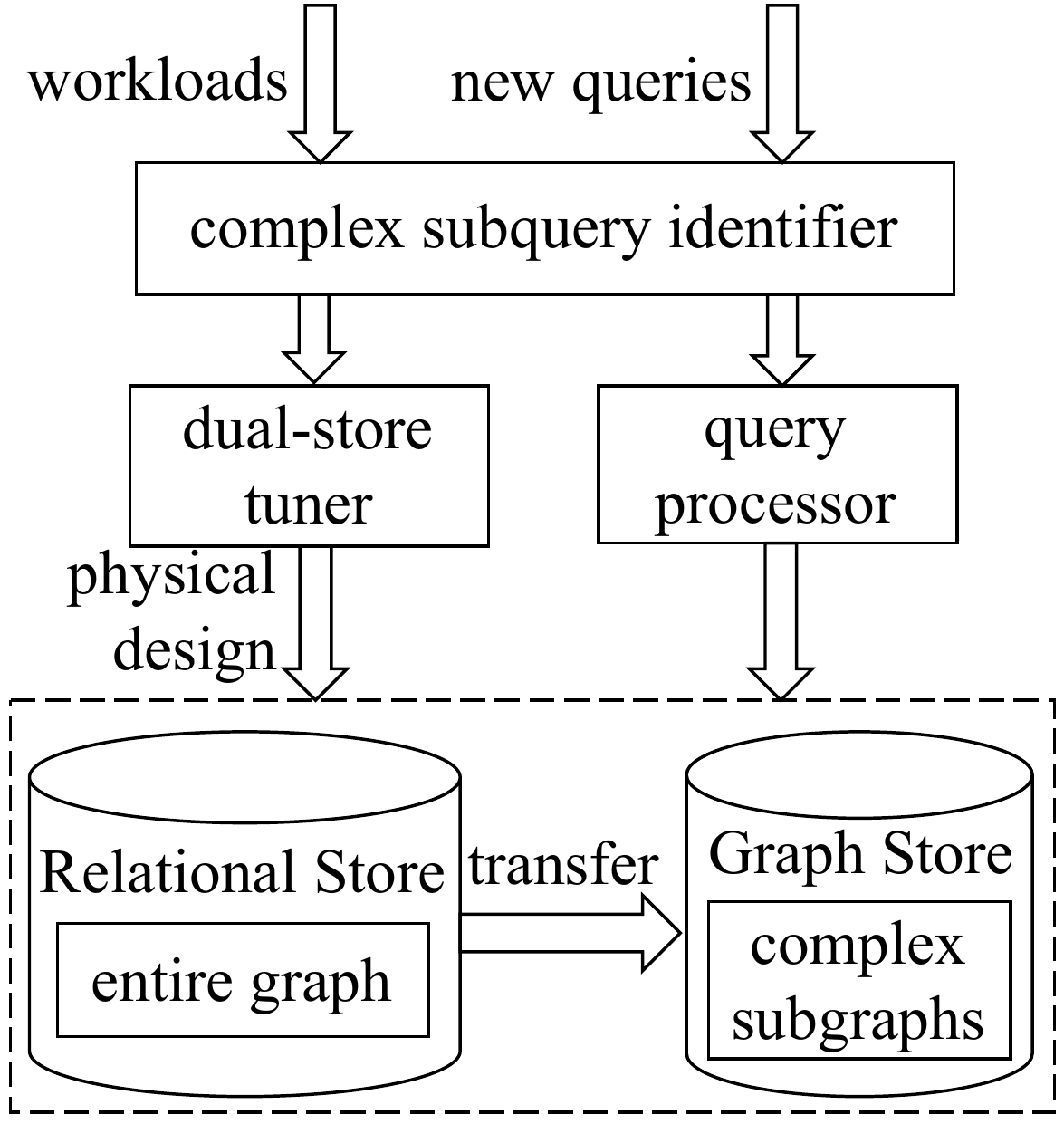}
	\caption{Dual-Store Structure for Knowledge Graphs}
	\label{fig:structure}
\end{figure}

Our dual-store structure for knowledge graphs is sketched in Figure~\ref{fig:structure}. Its design has two goals. One is to make full use of the advantages of both relation-based and graph stores. The other is to adjust itself automatically to the dynamically changing workloads.

Relational stores are able to store large-scale knowledge graphs and convenient in updating knowledge, but as shown in Table~\ref{table:pre-exp}, when the selectivity of a knowledge graph query is large, the query performance weakens obviously. Native graph stores are efficient in processing complex knowledge graph queries due to its index-free adjacent property~\cite{Rodriguez2012The}, but they are inapplicable to manage a large-scale knowledge graph due to their limited storage budgets or inflexible updating process~\cite{pokorny2017integrity}. Therefore, for the first goal, we use a native graph database as the accelerator for complex queries, while storing and updating the entire knowledge graph in a relational database.

Since heuristic tuning methods fail to memorize the historical experience and tune the physical design automatically, we leverage the memory ability and automation of machine learning to tune the dual-store structure. Thus, to achieve the second goal, we incorporate machine learning techniques into dual-store physical design tuning according to the dynamic workloads.

As Figure~\ref{fig:structure} shows, in order to accelerate the large-selectivity complex queries with native graph stores, we add a complex subquery identifier to judge whether a query contains a complex subquery. If there is a complex subquery in the query, the complex subquery identifier marks it. To support query processing based on current state of dual-store structure, we add a query processor to handle all the new queries including the marked complex queries. In order to guarantee the self-adaptivity of dual-store structure, we also add a dual-store tuner to learn the most recent workloads and tune the physical design. To achieve this, the complex subquery identifier sends the marked complex queries in the workloads to dual-store tuner. Based on the accumulated learning experience of historical workloads, the tuner automatically transfers valuable data from the relational store to the graph store. With the help of these components, our dual-store structure not only speeds up the process of complex queries with large selectivity, but also adapts to the frequent changes of data and workloads.

Based on above discussions, the dual-store structure has three components, complex subquery identifier, dual-store tuner, and query processor. In the remaining part of this section, we discuss these three components.

\subsection{Complex Subquery Identifier}
The goal of this component is to identify the complex subquery of each query. In this paper, a complex subquery is a set of subqueries whose subject variable and object variable both occur more than once in the query. Here, we give an example to explain the complex subquery.

\begin{myEg}
	\label{eg1}
	Consider a knowledge graph query $q$:
	\begin{framed}
		\noindent SELECT ?GivenName ?FamilyName WHERE\{\\
		$q_1$:  ?p y:hasGivenName ?GivenName.\\
		$q_2$: ?p y:hasFamilyName ?FamilyName.\\
		$q_3$: ?p y:wasBornIn ?city.\\
		$q_4$: ?p y:hasAcademicAdvisor ?a.\\
		$q_5$: ?a y:wasBornIn ?city.\\
		$q_6$: ?p y:isMarriedTo ?p2.\\
		$q_7$: ?p2 y:wasBornIn ?city.\}
	\end{framed}
	
Since each of the variables ?p, ?city, ?a, and ?p2 occurs more than once in $q$, we treat the subquery set of $q_3$, $q_4$, $q_5$, $q_6$, and $q_7$ as a complex subquery. Its output is the variable which joins it and the remaining part of the query, that is the subquery set of $q_1$ and $q_2$. Thus, the complex subquery $q_c$ of $q$ is:
	
\begin{framed}
		\noindent SELECT ?p WHERE\{\\
		$q_3$: ?p y:wasBornIn ?city.\\
		$q_4$: ?p y:hasAcademicAdvisor ?a.\\
		$q_5$: ?a y:wasBornIn ?city.\\
		$q_6$: ?p y:isMarriedTo ?p2.\\
		$q_7$: ?p2 y:wasBornIn ?city.\}
\end{framed}
\end{myEg}

In the complex subquery identifier, when a query $q$ comes, we scan $q$ and extract the subqueries with each variable occurring more than once in $q$. The set of these subqueries is identified as the complex subquery $q_c$ of $q$. The time complexity of complex subquery identifier is $O$($n$), where $n$ is twice the number of subqueries in $q$.

\subsection{Dual-Store Tuner}
The aim of this component is to tune the physical design of dual-store structure according to the changing workloads. The dual-store tuner is invoked periodically to decide which triple partitions to transfer from the relational store to the graph store. In this paper, triple partition refers to a set of triples whose predicates are identical in a knowledge graph.

To determine which triple partitions are worthwhile transferring, we adopt Q-learning as the training model, and the historical complex subqueires from \emph{complex subquery identifier} as the training data, to train dual-store tuner. With the help of Q-learning, our tuner selects triple partitions which maximize the potential query cost improvements based on actual runtime feedbacks. When triple partitions are migrated during tuning, they are stored in the permanent graph storage and become a part of the physical design until the next tuning phase. The details of this component will be discussed in Section~\ref{sec:dotil}.

\subsection{Query Processor}
Given the query $q$ and its complex subquery $q_c$ identified by \emph{complex subquery identifier}, if any, as input, this component aims to process $q$ based on current state of our store. The process of $q$ may rely on a single store, or span the dual stores and utilize the physical design of each store. This depends on the existing complex subgraphs in the native graph store.

If $q$ is processed in a single store, the query processor forwards it to the appropriate store. If the query process of $q$ spans both stores, the query processor first sends $q_c$ to the graph store. Then, it migrates the intermediate results from the graph store to the relational store. Finally, it resumes the remaining process of $q$ in the relational store. When the intermediate results are migrated, they are stored in the temporary relational table space, and discarded at the end of query process. We will discuss the details of this component in Section~\ref{sec:plan}.

\section{Dual-Store Tuner}
\label{sec:dotil}
In order to accelerate the complex query process, we tune the physical design of our dual-store structure according to the dynamic workloads. In this section, we describe the details of dual-store tuner component shown in Figure~\ref{fig:structure}. Our main idea is to transfer the triple partitions which maximize the potential complex query cost improvements of future workloads from the relational store to the graph store. Since it is challenging to decide which triple partitions are worthwhile transferring, we model the dual-store physical design tuning problem as a Markov Decision Process and develop a reinforcement-learning-based solution to identify the valuable triple partitions and generate the tuning policy.

\subsection{Problem Definition}
Since the elements of our dual-store physical design tuning are triple partitions, we denote the triple partition set in relational store and graph store as $T_R$ and $T_G$, respectively. Due to the limited storage capability of a graph store~\cite{ghrab2016grad}, its constraint is denoted as $B_G$.

Let $D$ = $\langle T_R$, $T_G\rangle$ be a dual-store design. $D$ is a pair where the first component denotes the triple partitions in the relational store, and the second component represents the triple partitions in the graph store.

We denote the historical knowledge graph complex query workload as $Q_c$. Since each query $q_c$ in $Q_c$ has a latency, we define the query costs of $Q_c$ with the given dual-store design $D$ as:

\begin{equation}
\label{eq:cost}
  Cost(Q_c, D) = \sum_{q_c\in Q_c}Latency(q_c,D).
\end{equation}

Intuitively, we measure the workload costs as the total latency time of query processing in the workload. This is a reasonable metric, but other metrics which measure the total expense of computational resources during the query process could also be applied such as total throughput.

Based on the definitions discussed above, we define the dual-store physical design tuning problem as follows.

\begin{myDef}
	(Dual-Store Physical Design Tuning Problem.) Given a batch of historical knowledge graph complex subqueries $Q_c$, a dual-store design $D$ = $\langle T_{R}$, $T_{G}\rangle$, and the storage constraint of graph store $B_{G}$, the dual-store design problem computes a new dual-store design $D^{new}$ = $\langle T_R^{new}$, $T_G^{new}\rangle$ that minimizes the query costs $Cost$($Q_c$, $D^{new}$) and satisfies the constraint $B_G$.
\end{myDef}

As defined above, the dual-store physical design tuning problem is a variant of knapsack problem~\cite{horowitz1974computing}. We can treat each triple partition in $T_R$ as an item, its transferring benefit as the weight of each item, the graph store as the knapsack, and its storage constraint $B_G$ as the knapsack capacity. Since the benefit of each triple partition is unknown and changing, the dual-store physical design tuning problem is more difficult than the knapsack problem. Therefore, the physical design tuning problem of dual-store knowledge graph structure is at least NP-hard.

\subsection{Our solution}
Since we desire a practical solution to the dual-store physical design tuning problem, we consider the actual situation of knowledge graph management. In practice, a knowledge graph manager suspends access to the graph, updates new knowledge, reconfigures related information, and resumes the access service periodically. The reconfiguration includes physical design tuning of knowledge graph store. Since the tuning process is invoked periodically, we adjust our dual-store structure according to the most recent batch of knowledge graph queries, which reflects the features of future workloads best. Moreover, the offline property makes the complex query process high-efficiency without any influence of physical design tuning.

Motivated by these, we first model the dual-store physical design tuning problem as a Markov Decision Process, and then train a reinforcement-learning-based tuner during each period of offline tuning with the most recent batch of workload as the training data. Assume there are $n$ triple partitions, the state space is $2^n$. Thus, the new challenge is how to train the reinforcement learning model effectively with such a large state space and limited workloads. To tackle this challenge, we consider a state space decomposition strategy to increase the retraining frequency of each state. In addition, we incorporate a counterfactual scenario into our tuner to guarantee the utilization of each complex query in the workloads.

\subsubsection{Modeling}
The goal of our tuning process is to determine which triple partitions are worthwhile transferring from the relational store to the graph store. Since each tuning operation is affected by the current state of dual-store structure as the results of last operation, we attempt to model our physical design tuning problem as a Markov Decision Process.

First, we denote the triple partition set in a knowledge graph as $T$ = \{$T_1$, $T_2$, ..., $T_n$\}, where $n$ is the number of predicates in the knowledge graph. Initially, we store the entire set $T$ in the relational store, and the graph store is empty. If $T_i$ (1$\leq$i$\leq$n) is in the graph store, we denote its status as 1. Otherwise, its status is 0. Therefore, the initial state of $T$ is \{0, 0, ..., 0\} and the state space of each $T_i$ is \{0, 1\}. In our tuning process, we decide which $T_i$ transfers to the graph store at which time. When the storage size achieves the constraint of graph store, we determine to evict which  $T_i$ from graph store. The action of transferring or evicting $T_i$ is denoted as 1, while keeping $T_i$ is represented as 0. Thus, the action space of each $T_i$ is \{0, 1\}. Note that whether $T_i$ is stored in the graph store, it is not evicted from the relational store due to the efficiency of inserting, deleting, or updating knowledge graph data. Therefore, in our dual-store design $D$, $T_R$ remains the same unless data changes, $T_G$ is the set of $T_i$ whose value is 1.

Consider the complex query $q_c$ in Example~\ref{eg1}. We judge whether the triple partitions corresponding to ``y:wasBornIn'', ``y:hasAcademicAdvisor'', and ``y:isMarriedTo'', denoted as $T_1$, $T_2$, and $T_3$, need to transfer together since $q_c$ is able to be processed in the graph store only when all of $T_1$, $T_2$, and $T_3$ store there. If we decide to transfer them, a reward of this action occurs after $q_c$ is processed. The reward is cost difference of $q_c$ between in the last state of our dual-store design and in the current state, which is defined as:

\begin{equation}
\label{eq:reward}
	Reward(q_c) = Cost(q_c, D)-Cost(q_c, D_{new}).
\end{equation}

In Example~\ref{eg1}, ``y:wasBorn'' accounts for $\frac{3}{5}$ in $q_c$, while ``y:hasAcademicAdvisor'' and ``y:isMarriedTo'' each makes up $\frac{1}{5}$. If we assign the whole reward of $q_c$ by the contribution proportion of each $T_i$, the amortized reward of $T_1$ is $\frac{3}{5}Reward(q_c)$, the reward of $T_2$ is $\frac{1}{5}Reward(q_c)$, and that of $T_3$ is $\frac{1}{5}Reward(q_c)$. Therefore, the reward of $q_c$ can be divided into the amortized reward of each $T_i$.

\textbf{Decomposition Strategy.} Since the size of state space $T$ is 2$^n$, where $n$ is the amount of predicates in knowledge graph, it requires a large amount of training data. However, in practice, the possibility of occurring two identical states is very low. In order to achieve a good performance with limited training data, we decompose the state space $T$ into a set of state subspaces with each subspace in it denoted by $T_i$. Thus, the combinations of all subspaces are able to represent all the possible states of $T$. For each $T_i$, the state space is \{0, 1\}, where 0 represents that $T_i$ is not stored in the graph store, 1 denotes $T_i$ stores there. The action space is \{0, 1\}, where 0 represents keeping $T_i$ in the current state, 1 denotes transferring $T_i$ from the relational store to the graph store or evicting $T_i$ from the graph store. $R$(0, 0) represents the reward of keeping $T_i$ in the relational store, which is kept 0. $R$(0, 1) denotes the amortized reward of transferring $T_i$ from the relational store to the graph store. $R$(1, 0) represents the accumulated reward of keeping $T_i$ in the graph store from its recent migration. $R$(1, 1) denotes the reward of evicting $T_i$ from the graph store, which is kept 0.

The reward of each action reflects the benefit or harm of keeping, transferring or evicting a triple partition. Our objective is to find a policy, that is a sequence of actions, to optimize the cumulative reward of workload which is defined as:

\begin{equation}
\label{eq:cumulative}
Reward(Q_c) = \sum_{q_c\in Q_c}Reward(q_c).
\end{equation}

To summarize, we model the dual-store physical design tuning problem as a Markov Decision Process as follows.

\begin{framed}
\noindent\textbf{State}: \{0, 1\} for each $T_i$ in $T$ \\
\textbf{Action}: \{0, 1\} keep, transfer, or evict $T_i$ \\
\textbf{Reward}: $R$(0, 0) and $R$(1, 0) are the rewards of keeping $T_i$, $R$(0, 1) is the reward of transferring $T_i$, $R$(1, 1) is the reward of evicting $T_i$ \\
\textbf{Policy}: decision to keep, transfer, or evict $T_i$
\end{framed}

\subsubsection{DOTIL}
Naturally, we adopt reinforcement learning to optimize the Markov Decision Process objective and propose a Dual-stOre Tuner based on reInforcement Learning, named \emph{DOTIL}. Since the state space in our setting is relative small, to make the training process lightweight, we select Q-learning as the reinforcement learning algorithm.

\textbf{Policy Update.} Since our dual-store tuner works periodically, we train the tuning policy with the most recent batch of knowledge graph queries based on Q-learning. The Q-function is defined as:
\begin{equation}
\label{eq:ql}
	Q^{new}(s_t, a_t)\leftarrow (1-\alpha)\cdot Q(s_t, a_t)+\alpha\cdot(r_t+\gamma\cdot\max_a Q(s_{t+1}, a))
\end{equation}
where $\max\limits_a Q(s_{t+1}, a)$ is the estimate of optimal future Q-value, $\gamma$ is the discount factor, $r_t$ is the reward of at time $t$, $\alpha$ is the learning rate, $\alpha\cdot(r_t+\gamma\cdot\max\limits_a Q(s_{t+1}, a))$ is the learned value, and $Q(s_t, a_t)$ is the old value.

For each $T_i$, there is a 2$\times$2 Q-matrix. During the training process, we update the values, i.e. $Q$(0, 0), $Q$(0, 1), $Q$(1, 0), $Q$(1, 1), in the matrix according to Equation~\ref{eq:ql}. Based on the current status and Q values, we decide the next action and update the tuning policy.

\textbf{Counterfactual Scenario.} As shown in Equation~\ref{eq:reward}, the reward $r_t$ is the cost improvement of querying $q_c$ in the current state of our dual-store design rather than the last state. If the set of $T_i$ corresponding to $q_c$ is transferred to the graph store, $r_t$ is the cost improvement of querying $q_c$ in the graph store rather than the relational store. However, $q_c$ is only processed in the graph store in our setting. Therefore, we attempt to create a counterfactual scenario which is not actually happened to obtain the query cost of $q_c$ in the relational store. To achieve this, we add a parallel thread to process $q_c$ in the relational store and returns its query cost to the main thread. Since the query costs of complex query in the relational store is greatly higher than that in the graph store, a threshold $\lambda$ is set to constraint the ratio of cost in the relational store to that in the graph store. In this way, we obtain the cost improvement of $q_c$, i.e. $r_t$.

Before introducing the algorithm of dual-store tuner, we define several functions used through the paper in Table~\ref{table:func}.

\begin{table*}[!htb]
	\centering
	\caption{Function Definitions}
	\label{table:func}
	\begin{tabular}{cc}
		\hline
		\textbf{Function} & \textbf{Definition} \\
		\hline
		\textsf{getPredicateSet()} & Given a query $q$, a graph $G$, or a triple partition set $T$, returns the set of predicates contain in $q$, $G$, or $T$. \\
		\textsf{relationQuery($q$)} & Given a query $q$, denotes processing $q$ in a relational store. \\
		\textsf{graphQuery($q$)} & Given a query $q$, denotes processing $q$ in a graph store.  \\
		\textsf{getResultSet()} & Given a query plan $P$, returns the result set of executing $P$. \\
		\textsf{migrate($S$, $s_1$, $s_2$)} & Given a query result set or a triple partition set $S$, two data stores $s_1$ and $s_2$, denotes migrating $S$ from $s_1$ to $s_2$. \\
		\textsf{getPartition()} & Given a predicate set $P$, returns the corresponding triple partition $T$. \\
		\textsf{getQmatrix()} & Given a triple partition $T_i$, returns the corresponding Q matrix of $T_i$. \\
		\textsf{getProportion($q$)} & Given a predicate $P_i$, a query $q$, returns the proportion of $P_i$ in the predicates contained in $q$.\\
		\textsf{stop($proc$)}  & Given a running query process $proc$, stops $proc$. \\
		\textsf{getCost()}  & Given a query process $proc$, returns the cost of $proc$. \\
		\textsf{getExectime()} & Given a running query process $proc$, returns the current execution time of $proc$. \\
		\textsf{getSize()} & Given a triple partition set $T$, returns the size of $T$. \\
		\textsf{des\_sort($obj$)}  & Given a triple partition set $T$, the value $v_i$ of an object $obj$ related to each $T_i$ in $T$, sorts $T_i$ according to the descending order of $v_i$. \\
		\textsf{evict($T$, $s$)} & Given a triple partition set $T$ and a data store $s$, evict $T$ from $s$. \\
		\hline
	\end{tabular}
\end{table*}

The pseudo code of \emph{DOTIL} is shown in Algorithm~\ref{algo:dotil}. The input is the current dual-store physical design $D$ consisting of $T_R$ and $T_G$, the storage constraint of graph store $B_G$, the most recent batch of complex subqueries $Q_c$, the current $Q$-matrix of each $T_i$, learning rate $\alpha$, discount factor $\gamma$, and a threshold $\lambda$. The algorithm produces a new dual-store physical design $D^{new}$.

\begin{algorithm}
	\caption{\emph{DOTIL}}
	\label{algo:dotil}
	\KwIn{current dual-store physcial design $D$=$<$$T_R$, $T_G>$, storage constraint of graph store $B_G$, most recent batch of complex subqueries $Q_c$, current $Q$-matrix of each $T_i$, learning rate $\alpha$, discount factor $\gamma$, threshold $\lambda$}
	\KwOut{new dual-store physical design $D^{new}$=$<T_R^{new}$, $T_G^{new}>$}
	$T_R^{new}\leftarrow T_R$, $T_G^{new}\leftarrow T_G$ \;
	\ForEach{$q_c\in Q_c$}
	{
		$P_c\leftarrow q_c.$\textsf{getPredicateSet()} \;
		$T_c\leftarrow P_c.$\textsf{getPartition()} \;
		\eIf{$T_c\subseteq T_G$}{\textsf{LearningProc($q_c$, $T_c$, 1, 0, $\alpha$, $\gamma$, $\lambda$)} \;
			\textbf{continue} \; }
		{
			$T_{set}\leftarrow\emptyset$, $sum_1\leftarrow$0, $sum_2\leftarrow$0 \;
			\ForEach{$T_i\in T_c$ $\&$ $T_i\notin T_G$}{$T_{set}.$add($T_i$)}
			\ForEach{$T_i\in T_{set}$}{
				$Q_i\leftarrow T_i.$\textsf{getQmatrix()} \;
				$Q_{00}\leftarrow sum_1+Q_i[0, 0]$ \;
				$Q_{01}\leftarrow sum_2+Q_i[0, 1]$ \;
			}
			\If{$Q_{00}\geq Q_{01}$}{\textbf{continue} \;}
			\If{$T_{set}.$\textsf{getSize()}+$T_G.$\textsf{getSize()}$>B_G$}{
				\ForEach{$T_i\in T_G$}{
					$Q_i\leftarrow T_i.$\textsf{getQmatrix()} \;
					$T_G\leftarrow T_G.$\textsf{des\_sort($Q_i[1, 1]$-$Q_i[1, 0]$)} \;
				}
				$rs\leftarrow Tset$.\textsf{getSize()}+$T_G.$\textsf{getSize()}-$B_G$ \;
				\ForEach{$T_i\in T_G$}{
					\textsf{evict($T_i$, graphStore)} \;
					$T_G^{new}\leftarrow T_G^{new}\setminus T_i$ \;
					\If{$r_s+T_i.$\textsf{getSize()}$\geq$0}{\textbf{continue} \;}
				}
			}
			\textsf{migrate($T_{set}$, relStore, graphStore)} \;
			$T_G^{new}\leftarrow T_G^{new}.$\textsf{add($T_{set}$)} \;
			\textsf{LearningProc($q_c$, $T_{set}$, 0, 1, $\alpha$, $\gamma$, $\lambda$)} \;
			\textsf{LearningProc($q_c$, $T_c\setminus T_{set}$, 1, 0, $\alpha$, $\gamma$, $\lambda$)} \;
		}
	}
	$D^{new}\leftarrow<T_R^{new}, T_G^{new}>$ \;
	\Return $D^{new}$\;
\end{algorithm}

We tune the physical design $D$ based on each complex subquery $q_c$ in $Q_c$ (Lines 1 and 2). Initially, we obtain the triple partition set $T_c$ corresponding to the predicate set $P_c$ in $q_c$ (Lines 3 and 4). If the triple partitions $T_G$ in graph store contain $T_c$, we train the Q-matrix $Q_i$ of each $T_i$ in $T_c$ (Lines 5-7). If $T_G$ does not contain $T_c$, we first obtain the set $T_{set}$ of triple partitions which are in $T_c$ but not in $T_G$ (Lines 8-11). Then, we compare the Q-value $Q_{00}$ of keeping $T_{set}$ in relational store and the Q-value $Q_{01}$ of transferring $T_{set}$ from relational store to graph store (Lines 12-15). If $Q_{01}$ is not greater than $Q_{00}$, we keep $T_{set}$ in the relational store (Lines 16 and 17). Otherwise, we judge whether the remaining size of graph store is enough to accommodate $T_{set}$. If it is not enough, we sort each triple partition $T_i$ in $T_G$ according to the difference between its Q-value of keeping in graph store and that of evicting from graph store (Lines 18-21). We evict $T_i$ in order until the remaining size is enough for $T_{set}$ and remove $T_i$ from $T_G^{new}$ (Lines 22-27). Then, we migrate $T_{set}$ from relational store to graph store and add $T_{set}$ to $T_G^{new}$ (Lines 28 and 29). Since $T_i$s in $T_{set}$ are transferred and those in $T_c$ but not in $T_{set}$ are kept in the graph store, we train the Q-matrix $Q_i$ of them respectively (Lines 30 and 31). Finally, we output $D^{new}$ consisting of $T_R^{new}$ and $T_G^{new}$ (Lines 32 and 33).

The pseudo code of function \textsf{LearningProc} is shown in Algorithm~\ref{algo:learning}. Next, we explain the Q-learning process of \emph{DOTIL} in detail.

The parameters of \textsf{LearningProc} include a query $q$, a triple partition set $T$, a state $s$, an action $a$, learning rate $\alpha$, discount factor $\gamma$, and a threshold $\lambda$. The function trains Q-matrix for each triple partition $T_i$ in $T$. First, we process query $q$ in the graph store and obtain its query cost $c_1$ (Line 1). Meanwhile, in a parallel thread, the query process $proc$ of $q$ in relational store is executed and monitored (Line 2). If the execution time of $proc$ achieves $\lambda$ times $c_1$, we stop $proc$ and obtain its current query cost $c_2$ (Lines 3-6). For each triple partition $T_i$ in $T$, we obtain the proportion of its corresponding predicate $P_i$ and compute its reward $r_t$ for the process of $q$ (Lines 7-10). With the given Q-matrix parameters $s$ and $a$, Q-learning parameters $\alpha$ and $\gamma$, we update the corresponding Q-value $Q_i[s,a]$ for each $T_i$ (Lines 11 and 12).

\begin{algorithm}
	\caption{\textsf{LearningProc($q$, $T$, $s$, $a$, $\alpha$, $\gamma$, $\lambda$)}}
	\label{algo:learning}
	\KwIn{query $q$, triple partition set $T$, state $s$, action $a$, learning rate $\alpha$, discount factor $\gamma$, threshold $\lambda$}
	\KwOut{learned Q-matrix}
	$c_1\leftarrow$\textsf{graphQuery($q$).getCost()} \;
	$proc\leftarrow$\textsf{relationQuery($q$)} \;  \tcp{parallel thread}
	\If{$proc.$\textsf{getExectime()}$=\lambda c_1$}{
		\textsf{stop($proc$)} \;
		$c_2\leftarrow\lambda c_1$ \;
	}
	$c_2\leftarrow proc.$\textsf{getCost()} \;
	\ForEach{$T_i\in T$}{
		$P_i\leftarrow T_i.$\textsf{getPredicateSet()} \;
 		$\delta(P_i)\leftarrow P_i.$\textsf{getProportion($q$)} \;
		$r_t\leftarrow(c_2-c_1)\times\delta(P_i)$ \;
		$Q_i\leftarrow T_i.$\textsf{getQmatrix()} \;
		$Q_i[s, a]\leftarrow(1-\alpha)\times Q_i[s, a]+\alpha\times(r_t+\gamma\times$ $\max\limits_aQ(s_{t+1},a))$ \;
	}
\end{algorithm}

The time complexity of Algorithm~\ref{algo:dotil} is determined by the number of knowledge graph complex queries $Q_c$ and the complexity of Algorithm~\ref{algo:learning}. Since the time complexity of \textsf{LearningProc} is positively related to the average cost $c_1$ of processing a complex query $q_c$ and the number of triple partitions $|T|$, the complexity of \emph{DOTIL} is affected by the values of $|Q_c|$, $c_1$, and $|T|$.

Here we explain two important issues to make our proposed \emph{DOTIL} applicable. First, since there is a cold start in Q-learning process, we prefer to warm up \emph{DOTIL} with historical queries. In this way, \emph{DOTIL} will achieve a better performance. Second, the initial Q-matrix $Q_i$ corresponding to each triple partition $T_i$ is a zero matrix. When the Q-value $Q_{00}$ of keeping $T_{set}$ in relational store and the Q-value $Q_{01}$ of transferring $T_{set}$ from relational store to graph store are both 0, we set an initial probability of transferring $prob$ to decide the first action of $T_i$. Since the value of $prob$ is positively related to the number of Q-learning model training, we prefer to set $prob$ not less than 50\%. We will discuss the detailed tuning of all the parameters in \emph{DOTIL} including $prob$ in Section~\ref{sec:exp}.

\section{Query Processor}
\label{sec:plan}
In this section, we focus on the query processor in the dual-store structure. The challenge of the query processor is how to efficiently process the given query based on the current dual-store state.

\begin{figure}
	\centering
	\includegraphics[width=3.2in,height=1.6in]{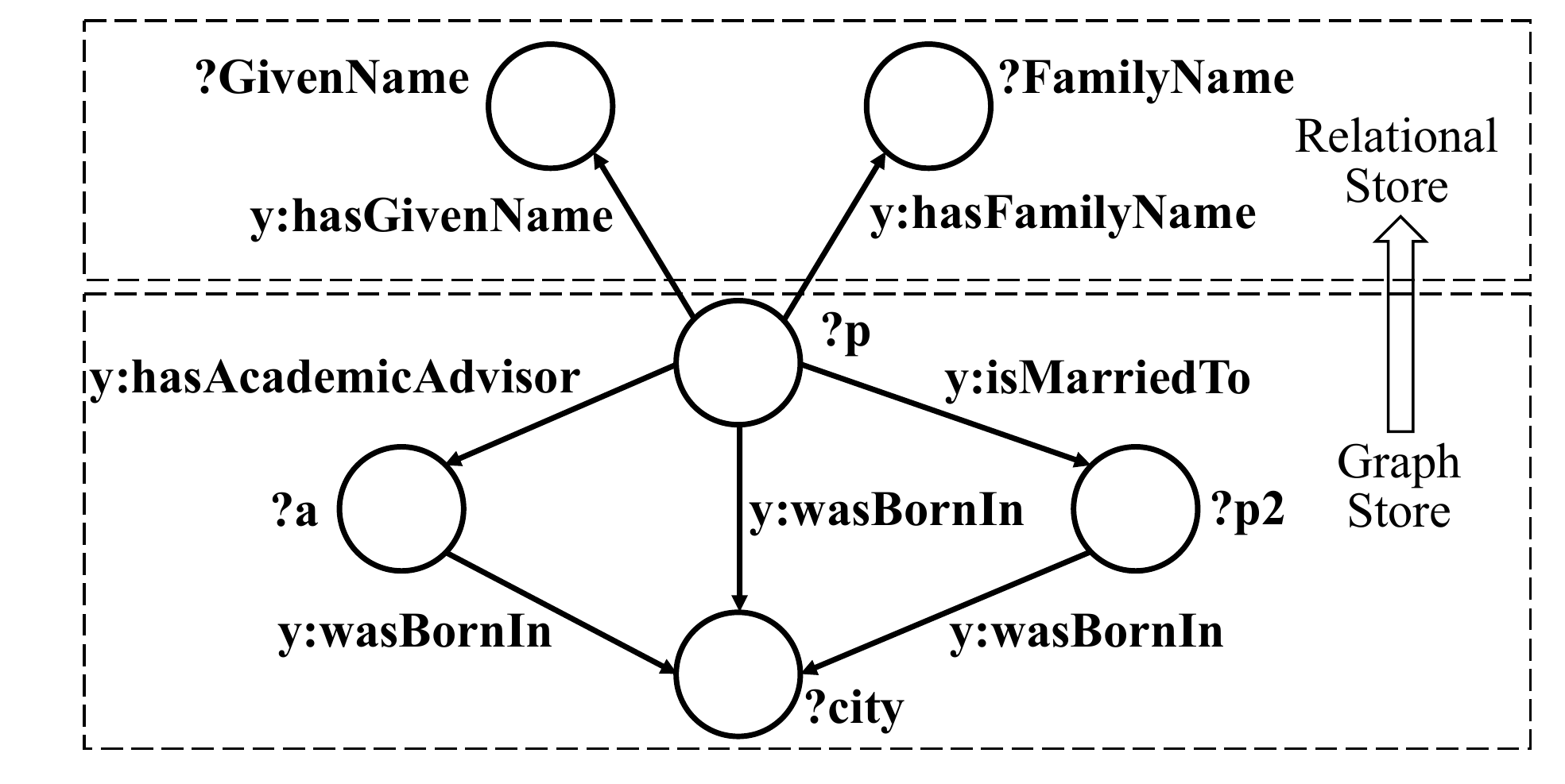}
	\caption{An Instance of Query Processing}
	\label{fig:split}
\end{figure}

In the light of this challenge, our main idea is that if the query contains a complex subquery, it is time-saving to process the subquery in a native graph store rather than a relational store due to the high cost of complex query process in relational databases. For example, we convert the query $q$ in Example~\ref{eg1} into a query graph. As depicted in Figure~\ref{fig:split}, a dual-store execution plan splits a query into two parts. The bottom part is its complex subquery $q_c$, which is to be executed in a graph store, while the remaining part of the query is processed in a relational store. Since the graph store is used as an accelerator, the intermediate results produced by it will be migrated to the relational store for further computation while the final query results are outputted directly.

However, whether a query $q$ or its complex subquery $q_c$ is able to be processed in a native graph store depends on the existing complex subgraphs. Thus, we consider the following three conditions.

\textbf{Case 1.} If the complex subgraphs in graph store cover all the predicates in $q$, we plan to process $q$ in the graph store.

\textbf{Case 2.} If the complex subgraphs in graph store do not cover all the predicates in $q$ but cover the predicates in $q_c$, we plan to process $q_c$ in the graph store, migrate the intermediate results from the graph store to the relational store, and finish the remaining part of $q$ in the relational store.

\textbf{Case 3.} If the complex subgraphs in graph store do not cover the predicates in $q_c$, we plan to process $q$ in the relational store.

The pseudo code of query processor is shown in Algorithm~\ref{algo:plan}. The input is a query $q$, its complex subquery $q_c$, and current complex subgraphs $G_c$ in the graph store. The algorithm processes the query $q$. We first judge whether $q$ contains a complex subquery $q_c$. If not, we plan to process $q$ in the relational store (Lines 1 and 2). If $q_c$ exists, we obtain the predicate sets of $q$, $q_c$, and $G_c$, respectively (Lines 3-6). If the predicates in $G_c$ cover the ones in $q$, we plan to process $q$ in the graph store (Lines 7 and 8). Otherwise, if the predicates in $q_c$ are covered by the ones in $G_c$, we plan to process $q_c$ in the graph store, migrate its result set from the graph store to the relational store, and process the remaining part of $q$ in the relational store (Lines 9-14). If the predicates in $q_c$ are not contained in the predicate set of $G_c$, we process $q$ in the relational store (Lines 15 and 16).

\begin{algorithm}
	\caption{Query Processor}
	\label{algo:plan}
	\KwIn{query $q$, complex subquery $q_c$ of $q$, current complex subgraphs $G_c$ in graph store}
	\eIf{$q_c$ is null}{\textsf{relationQuery($q$)}\;}
	{$P_q\leftarrow q.\textsf{getPredicateSet()}$\;
	$P_{q_c}\leftarrow q_c.\textsf{getPredicateSet()}$\;
	$P_{G_c} \leftarrow G_c.\textsf{getPredicateSet()}$\;
	\eIf{$P_q\subset P_{G_c}$}{\textsf{graphQuery($q$)}\;}
	{\eIf{$P_{q_c}\subset P_{G_c}$}
		{$res\leftarrow$ \textsf{graphQuery($q_c$).getResultSet()}\;
		 \textsf{graphQuery($q_c$)\;
		 migrate($res$, \textsf{graphStore}, \textsf{relStore})\;
		 relationQuery($q\backslash q_c$)$\rangle$} \;}
		{\textsf{relationQuery($q$)}\;}
	}	
	}
\end{algorithm}

Clearly, the time complexity of Algorithm~\ref{algo:plan} is $O$($m$), where $m$ is the edge number of complex subgraphs in the graph store. In practice, $m$ is about 20\% to 30\% of the edge number of an entire knowledge graph. Thus, the query processor actually takes little overhead.

\section{Evaluation}
\label{sec:exp}
In this section, we discuss our experimental evaluation of our proposed knowledge-graph dual-store structure. First, we describe our experimental methodology in Section~\ref{subsec:method}. Then in Section~\ref{subsec:variants}, we compare our dual-store structure with other store variants to highlight the benefits of our storage design in processing complex knowledge graph queries. In Section~\ref{subsec:dotil}, we test the performance results of our dual-store tuner \emph{DOTIL}. Lastly, we compare the behavior of \emph{DOTIL} with other tuning algorithms in Section~\ref{subsec:tuners}.

\subsection{Methodology}
\label{subsec:method}
In this section, we describe knowledge graphs, workloads, metrics, and setup used in our experiments.

\emph{Knowledge Graphs and Workloads.} We use three public knowledge graphs, YAGO~\cite{yago}, WatDiv~\cite{watdiv}, and Bio2RDF~\cite{biordf}. Knowledge graph information is shown in Table~\ref{table:graphinfo}. YAGO contains all the triples in YagoFacts and the triples whose predicate is ``hasGivenName'' or ``hasFamilyName''. We generate WatDiv by setting 14,634,621 as the number of triples. Bio2RDF consists of the knowledge from Interaction Reference Index, Online Mendelian Inheritance in Man, Pharmacogenomics Knowledge Base, and PubMed. We use the query templates of YAGO and Bio2RDF provided in~\cite{harbi2016accelerating}. The query templates of WatDiv are composed of linear queries (WatDiv-L), star queries (WatDiv-S), snowflake-shaped queries (WatDiv-F), and complex queries (WatDiv-C)~\cite{wattemp}. The query workloads of YAGO (20 queries), WatDiv-L (35 queries), WatDiv-S (25 queries), WatDiv-F (25 queries), WatDiv-C (15 queries), and Bio2RDF (25 queries) consist of the original query templates and four mutations for each query. Each workload has an ordered version and a random version. In the ordered version, each query template and its mutations are clustered. In the random version, we randomize all the query templates and their mutations in the workload. In this paper, we set each batch of queries as $\frac{1}{5}$ knowledge graph query workloads.

\begin{table}[!htb]
	\centering
	\caption{Knowledge Graph Information}
	\label{table:graphinfo}
	\begin{tabular}{cccccc}
		\hline
		\textbf{name} & \textbf{size} & \textbf{triples} & \#\textbf{-S$\cup$O} & \#\textbf{-P} & \#\textbf{-queries}  \\
		\hline
		YAGO & 679.3M & 16418085 & 5593541 & 39 &  20  \\
		WatDiv & 2.08G & 14634621 & 1396039 & 86 & 100   \\
		Bio2RDF & 7.64G & 60241165 & 8914390  & 161 & 25  \\
		\hline
	\end{tabular}
\end{table}

\emph{Metrics.} Our primary metric is time-to-insight, named \emph{TTI}~\cite{Michael2011Big}, which is the total elapsed time from a batch of workload submission to completion. \emph{TTI} evaluates the costs of our online query process. Also, we use the sum of Q-matrix of each triple partition to measure the offline training effect. The effectiveness of \emph{DOTIL} is positively related to the values in Q-matrix.

\emph{Setup.} All experiments are conducted on a server with 32$\times$Intel Xeon(R) Gold 6151 CPU@3.00GHz, 128G memory, and a 148G hard disk, under Ubuntu 18.04.1 LTS. The relational database used in this work is MySQL 5.7.30, and the graph database is Neo4j 3.5.14. All algorithms are implemented in Python.

\subsection{Comparisons of Store Variants}
\label{subsec:variants}
\begin{figure}[htb!]
	\centering
	\subfigure[ordered YAGO]{
		\includegraphics[width=1.65in,height=1.48in]{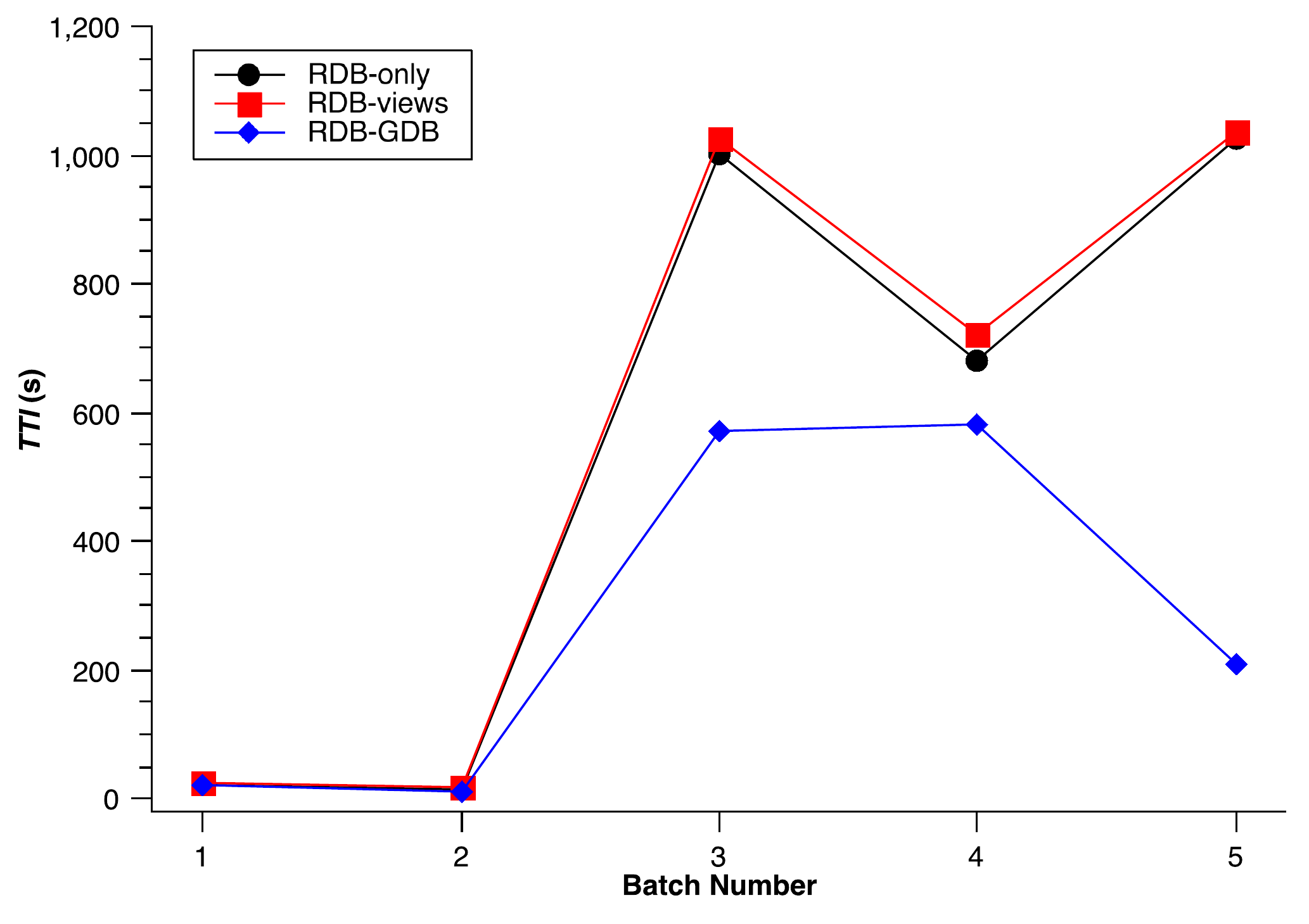}
		\label{fig:1-yago}
	}
	\subfigure[ordered WatDiv-L]{
		\includegraphics[width=1.65in,height=1.48in]{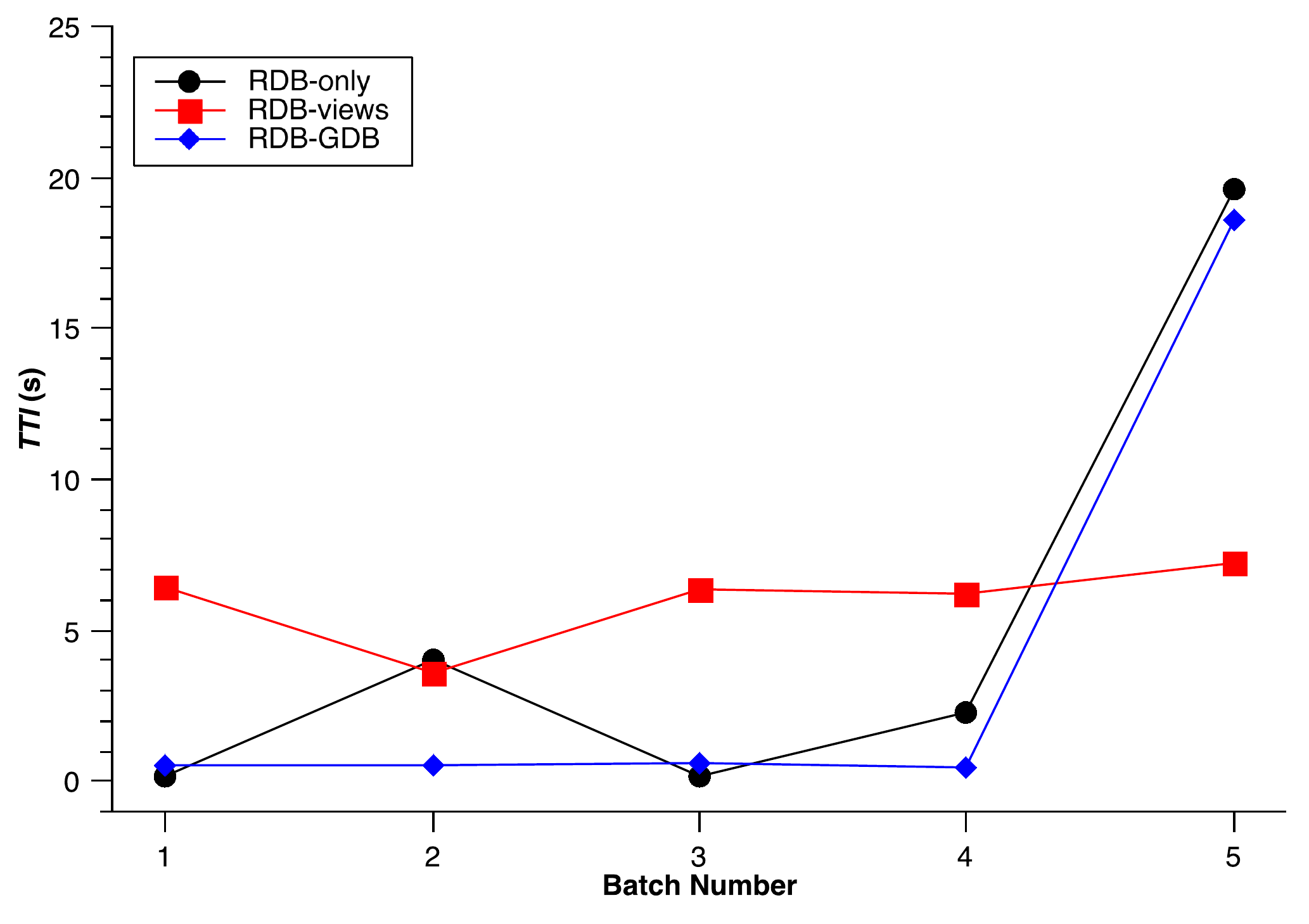}
		\label{fig:1-watl}
	}
	\subfigure[ordered WatDiv-S]{
	\includegraphics[width=1.65in,height=1.48in]{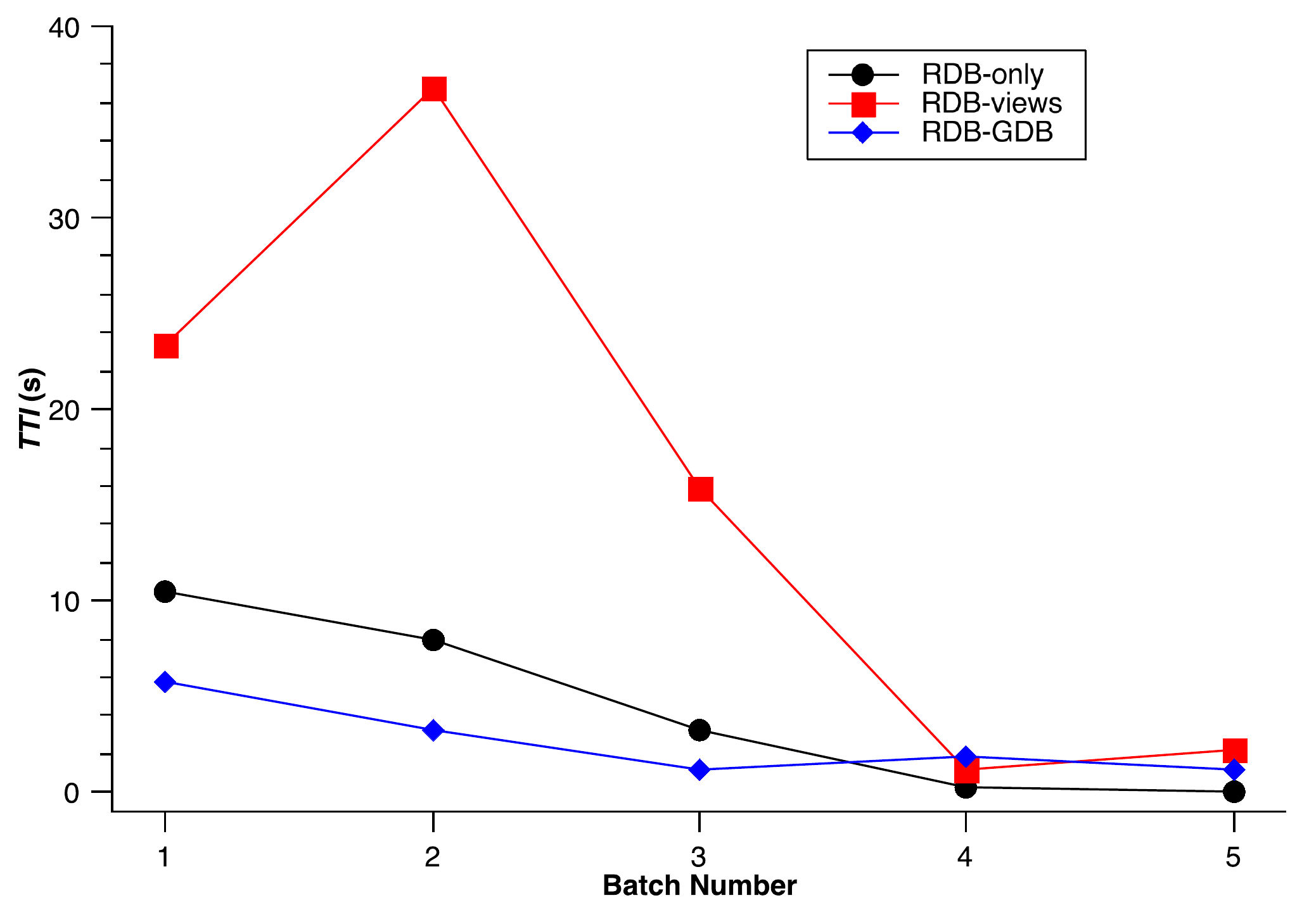}
	\label{fig:1-wats}
}
\subfigure[ordered WatDiv-F]{
	\includegraphics[width=1.65in,height=1.48in]{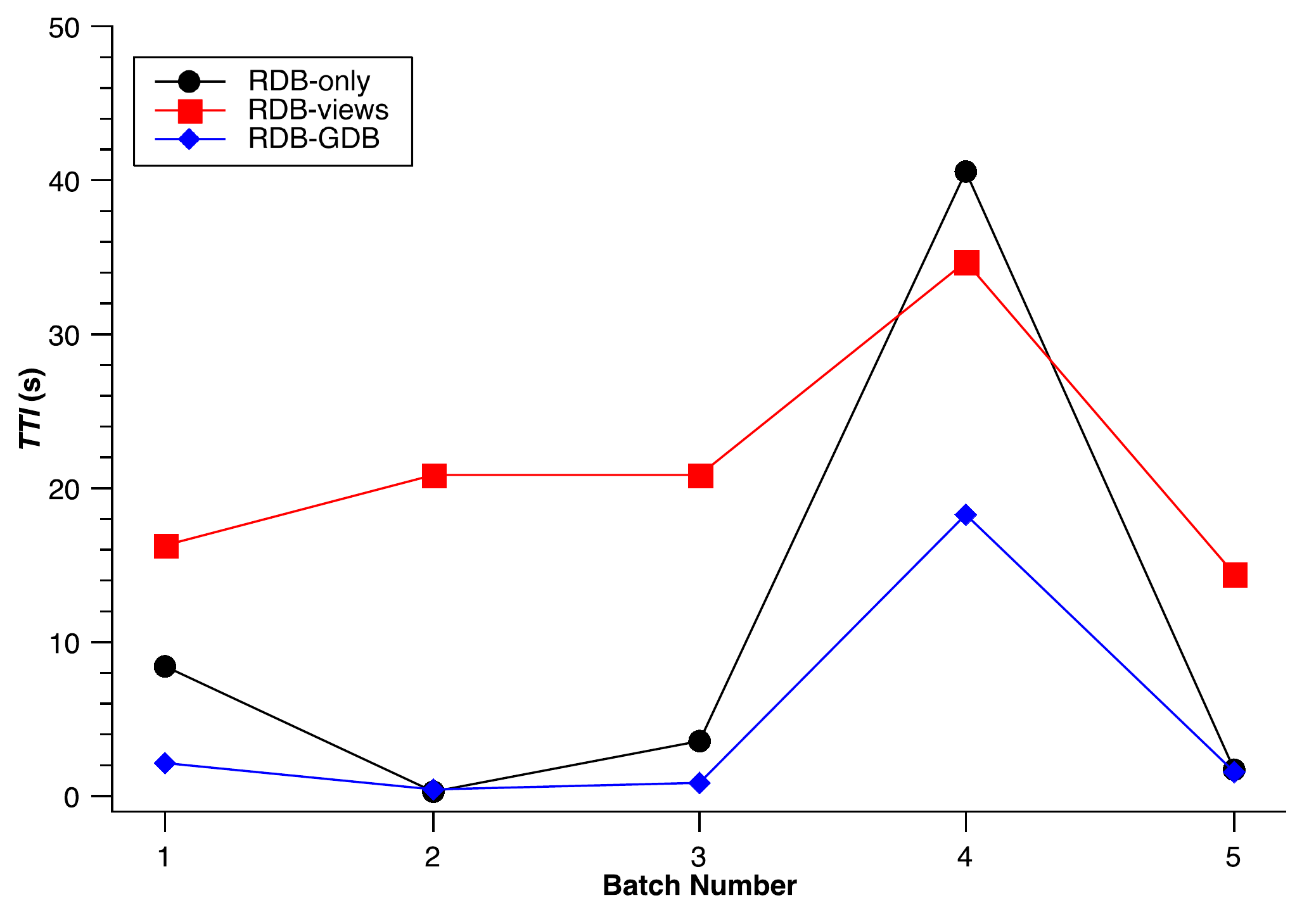}
	\label{fig:1-watf}
}
	\subfigure[ordered WatDiv-C]{
	\includegraphics[width=1.65in,height=1.48in]{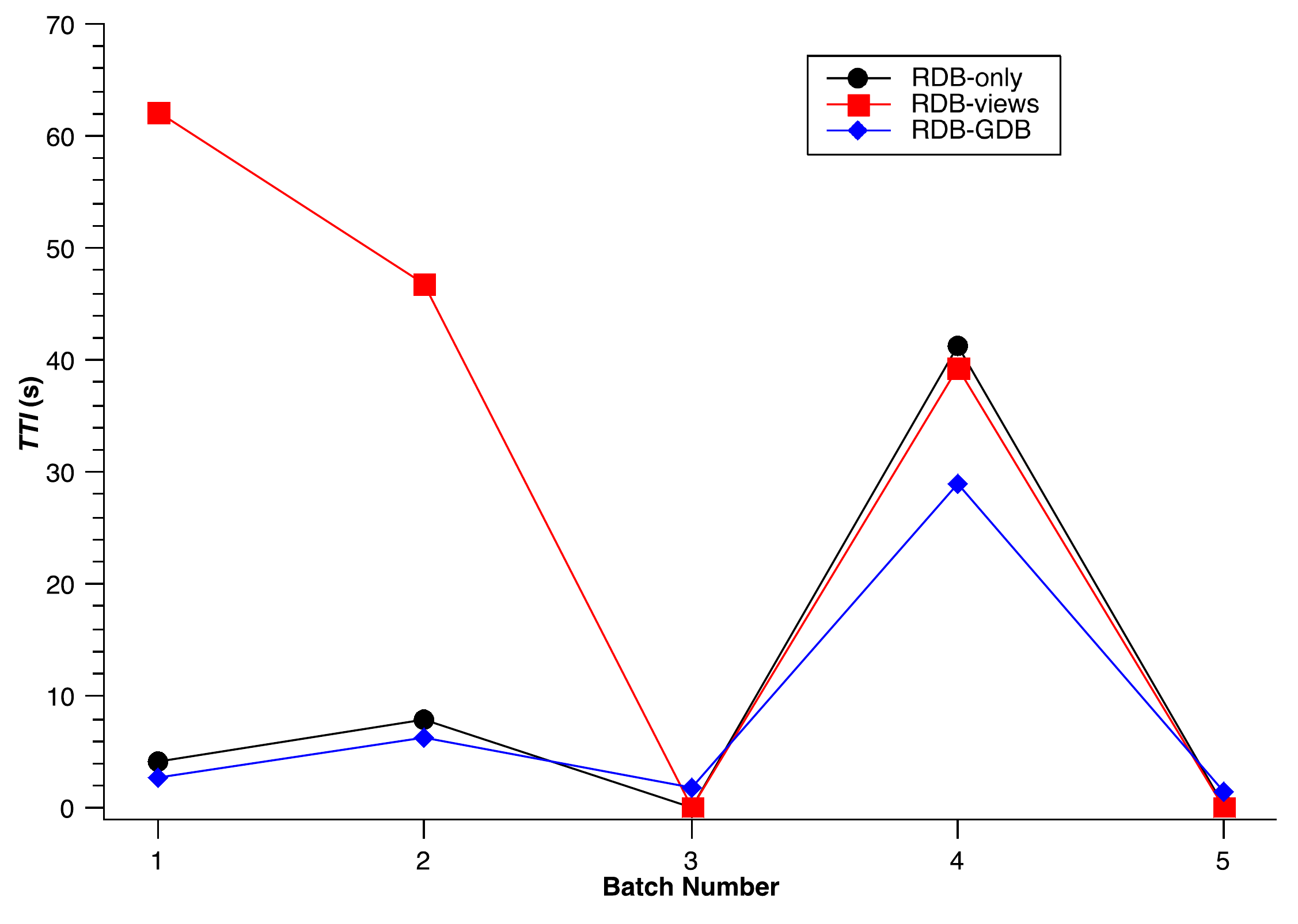}
	\label{fig:1-watc}
}
\subfigure[ordered Bio2RDF]{
	\includegraphics[width=1.65in,height=1.48in]{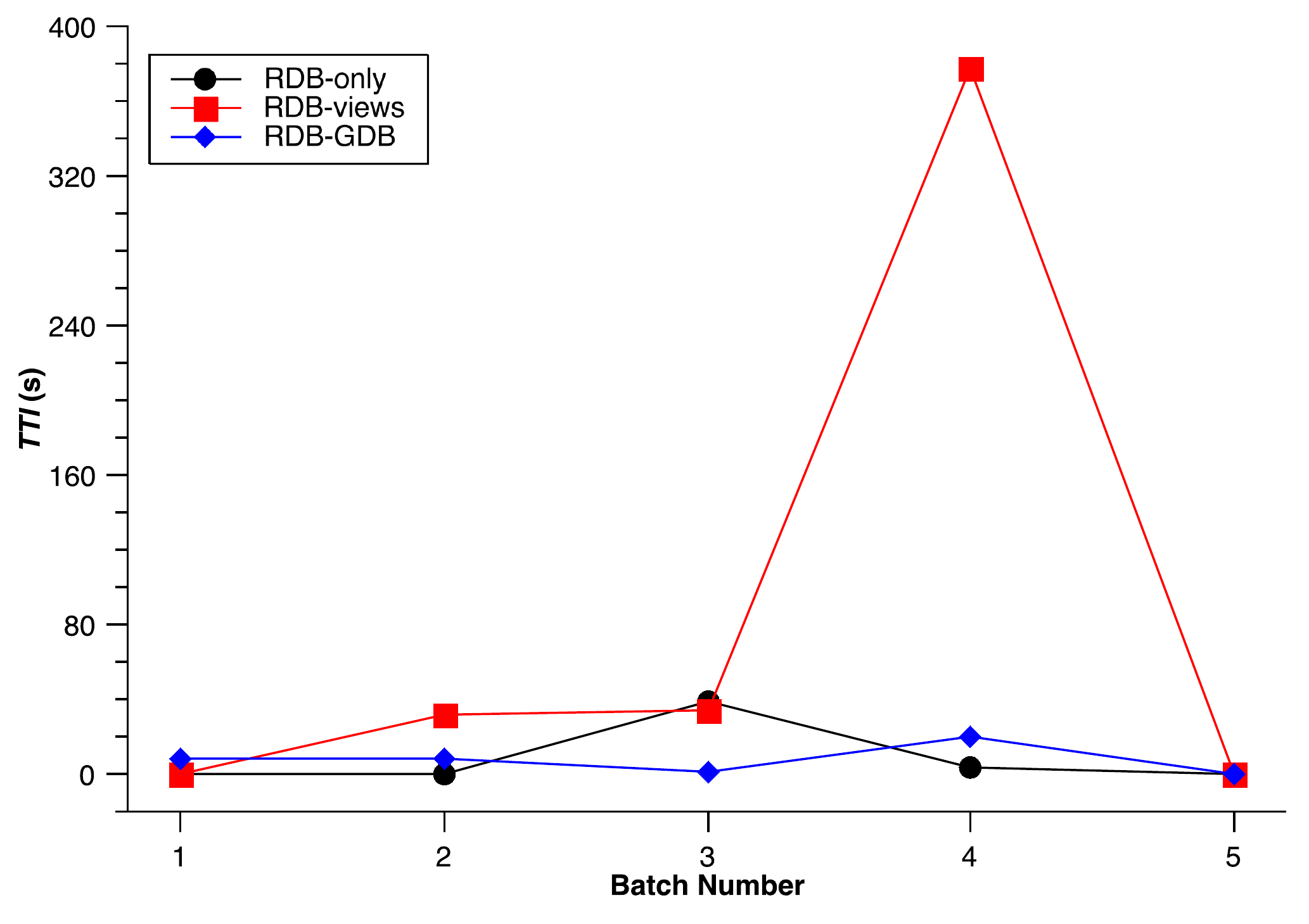}
	\label{fig:1-bio}
}
	\caption{Results of Each Batch Varying Store on Ordered Workloads}
	\label{fig:batch-store-order}
\end{figure}

\begin{figure}[htb!]
	\centering
	\subfigure[random YAGO]{
		\includegraphics[width=1.65in,height=1.48in]{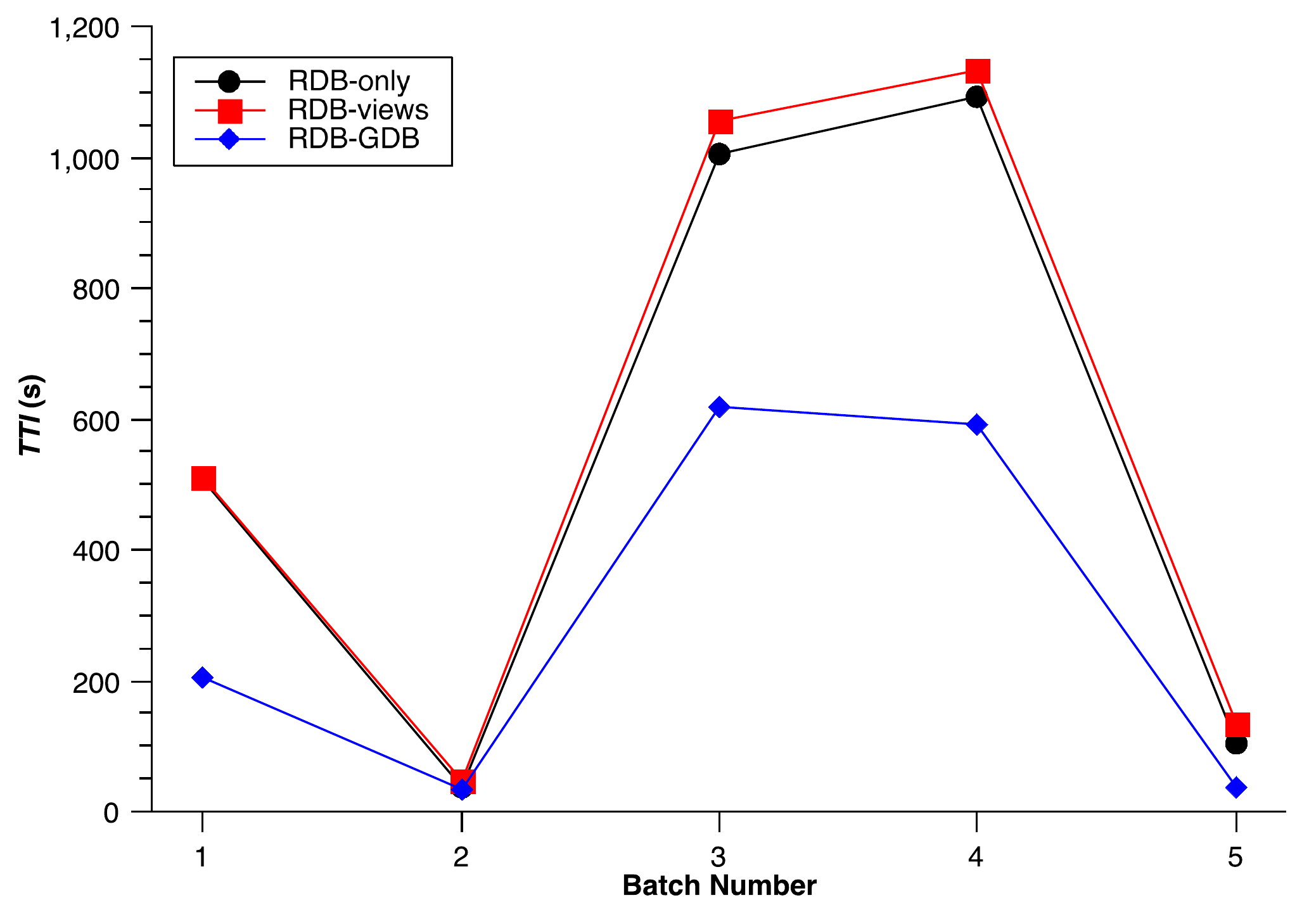}
		\label{fig:1-1-yago}
	}
	\subfigure[random WatDiv-L]{
		\includegraphics[width=1.65in,height=1.48in]{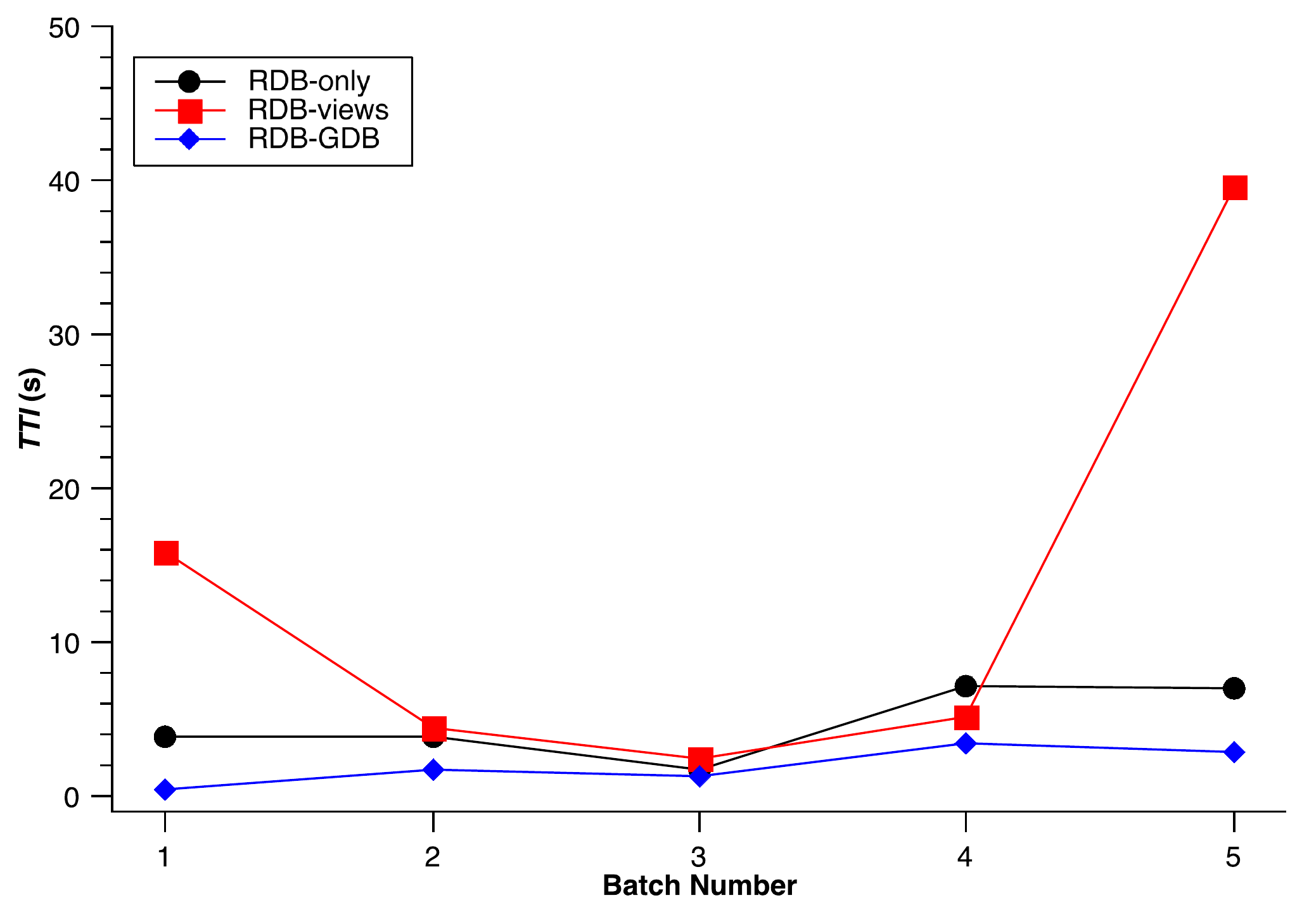}
		\label{fig:1-1-watl}
	}
	\subfigure[random WatDiv-S]{
		\includegraphics[width=1.65in,height=1.48in]{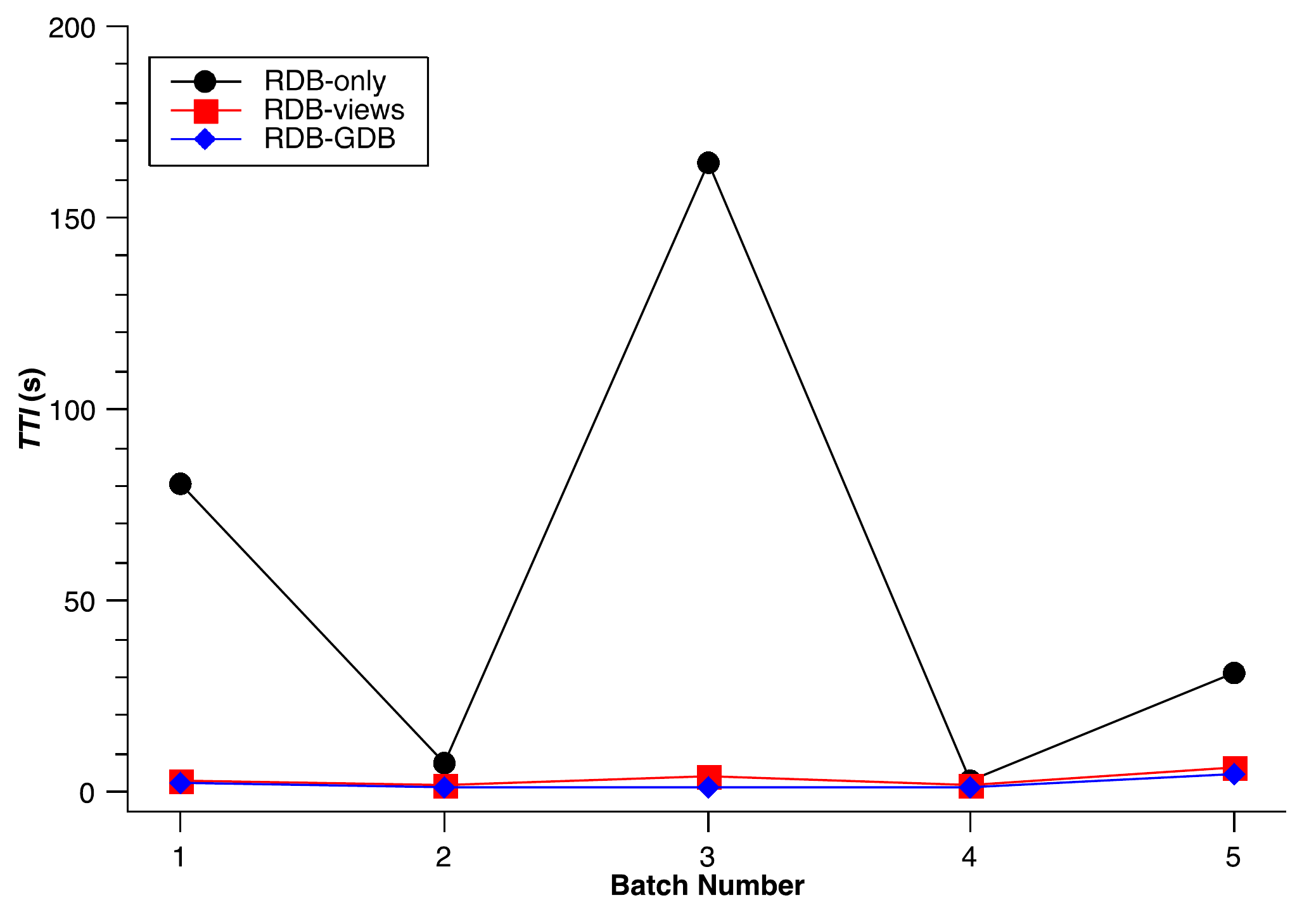}
		\label{fig:1-1-wats}
	}
	\subfigure[random WatDiv-F]{
		\includegraphics[width=1.65in,height=1.48in]{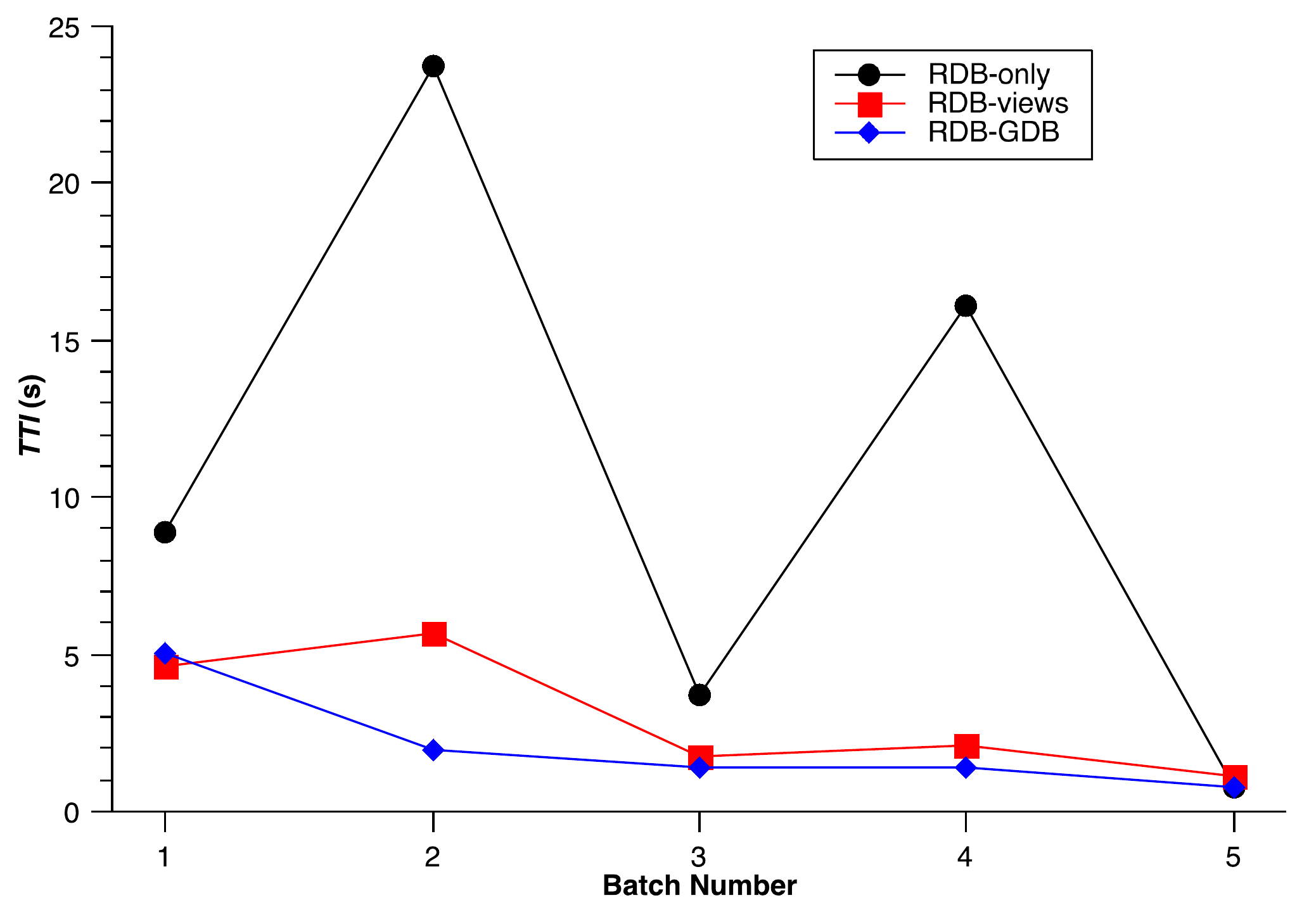}
		\label{fig:1-1-watf}
	}
	\subfigure[random WatDiv-C]{
		\includegraphics[width=1.65in,height=1.48in]{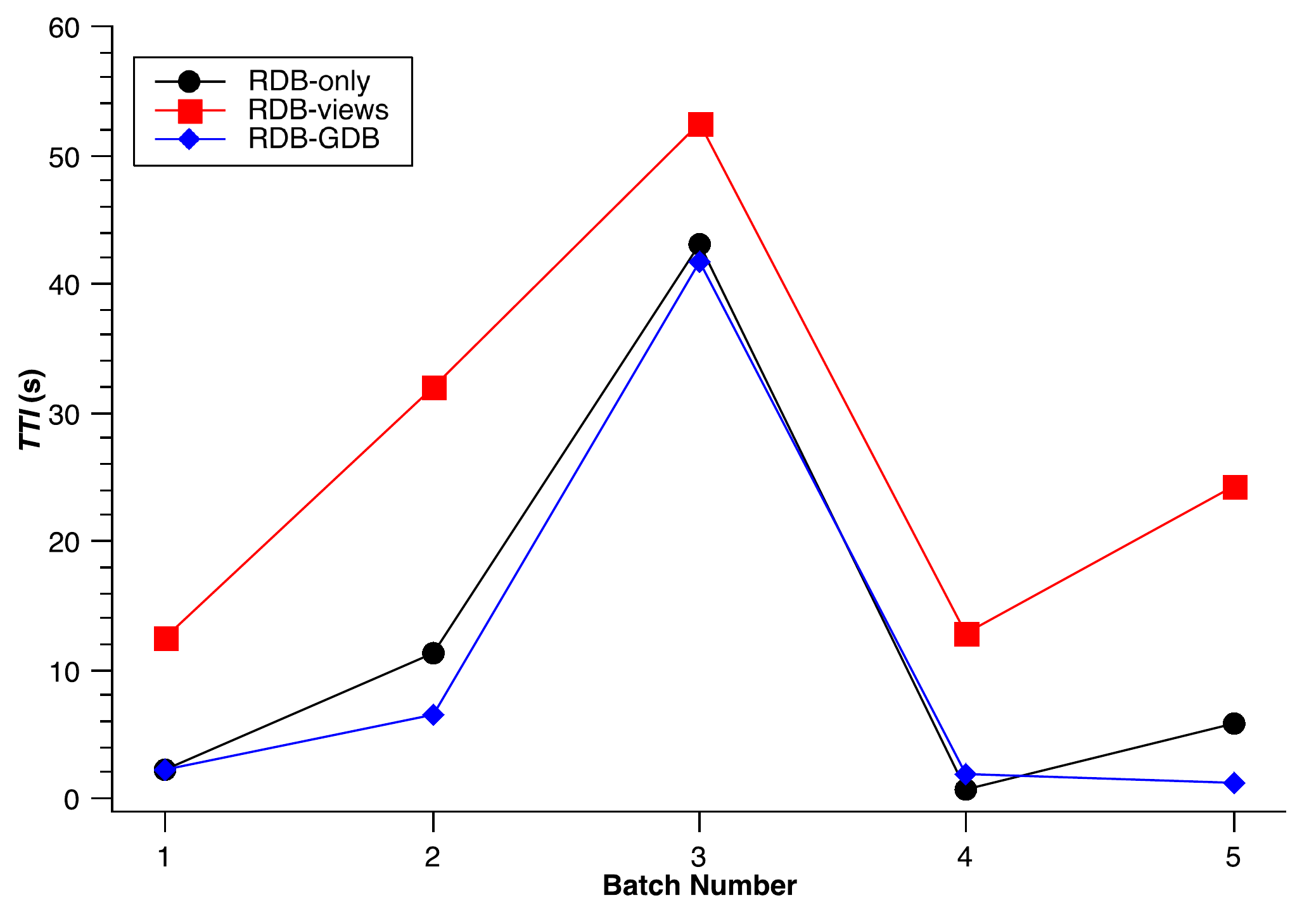}
		\label{fig:1-1-watc}
	}
	\subfigure[random Bio2RDF]{
		\includegraphics[width=1.65in,height=1.48in]{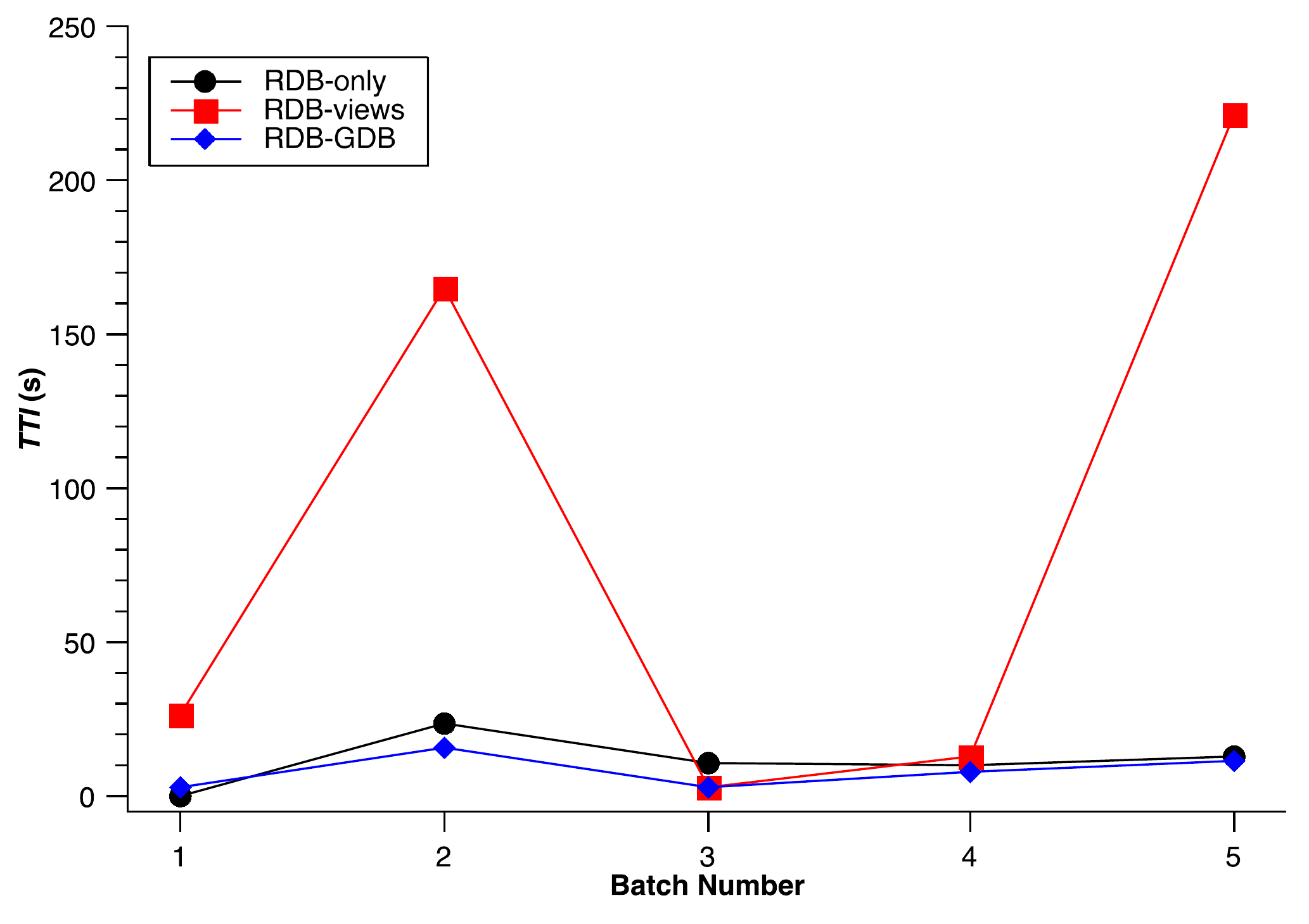}
		\label{fig:1-1-bio}
	}
	\caption{Results of Each Batch Varying Store on Random Workloads}
	\label{fig:batch-store-rand}
\end{figure}

\begin{figure}[htb!]
	\centering
	\subfigure[YAGO workloads]{
		\includegraphics[width=1.65in,height=1.48in]{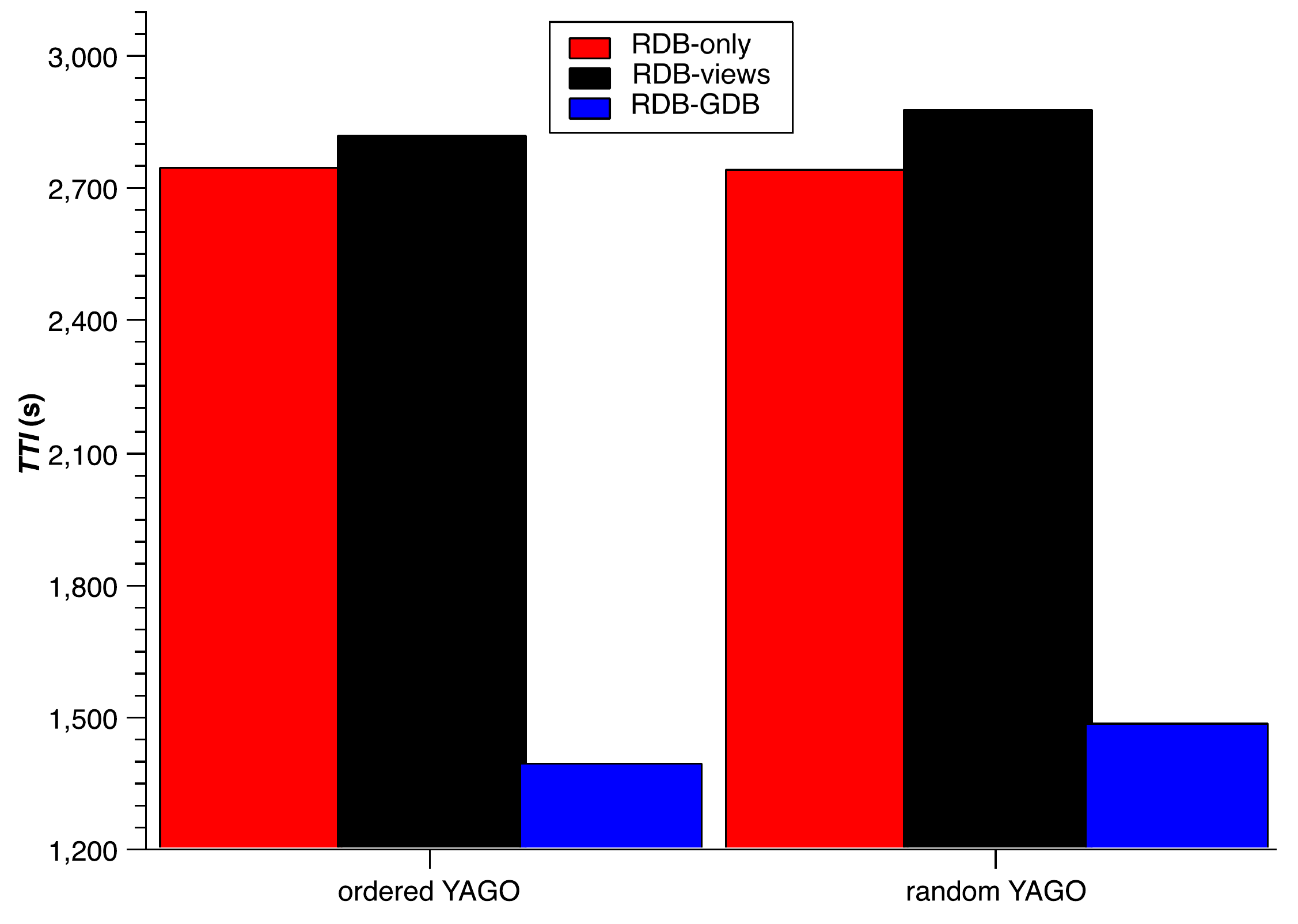}
		\label{fig:1-2-yago}
	}
	\subfigure[ordered WatDiv workloads]{
		\includegraphics[width=1.65in,height=1.48in]{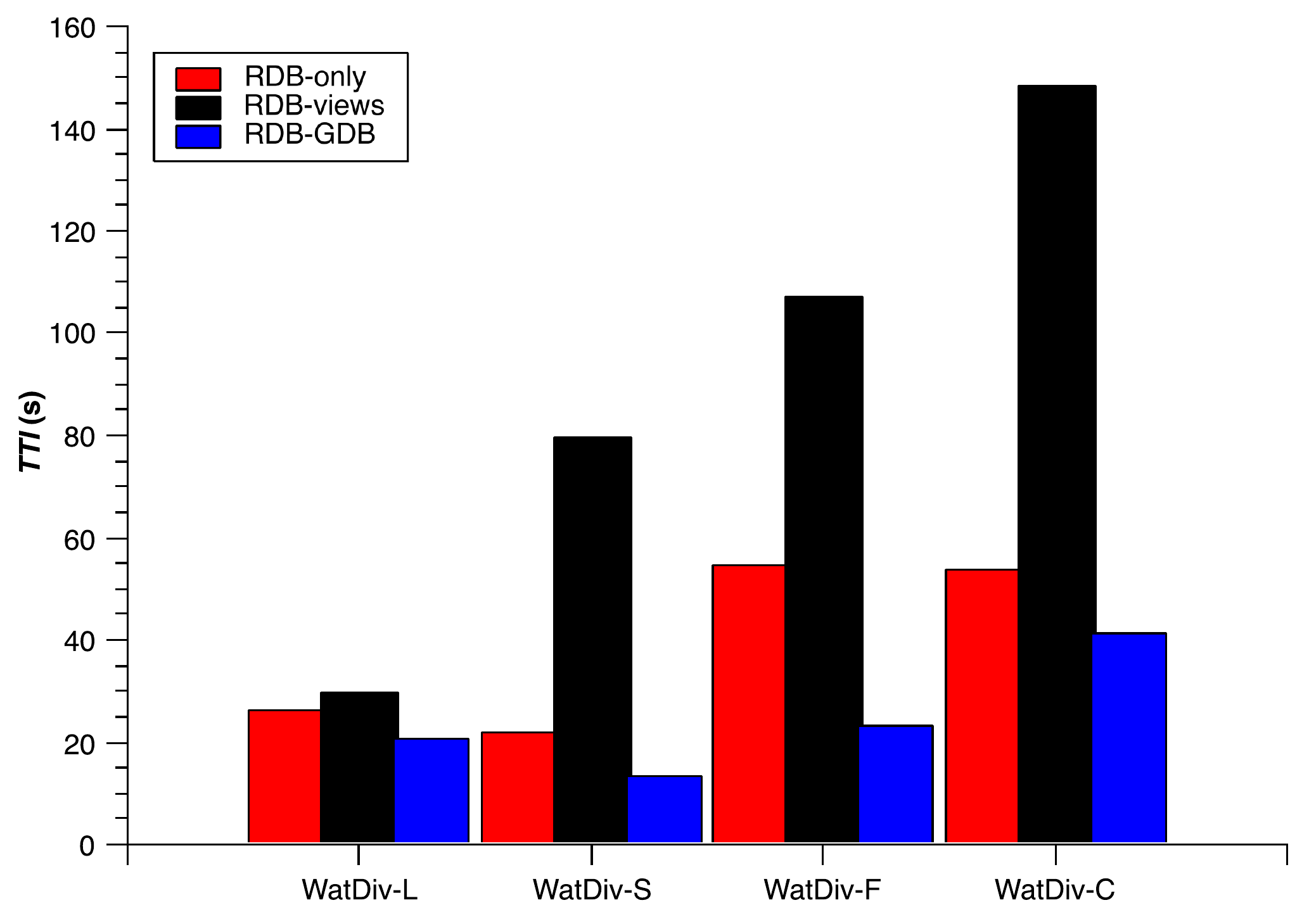}
		\label{fig:1-2-watorder}
	}
	\subfigure[random WatDiv workloads]{
		\includegraphics[width=1.65in,height=1.48in]{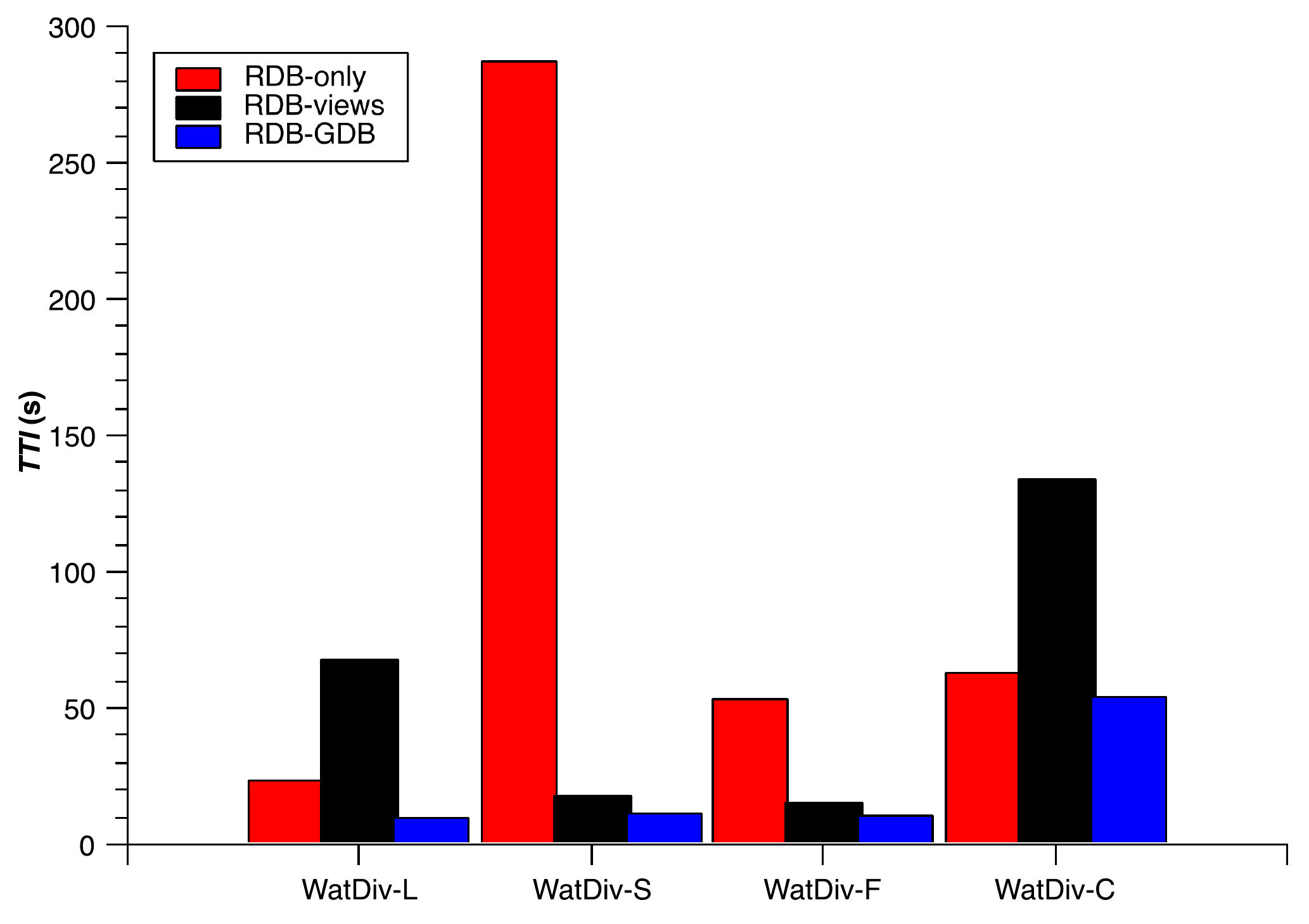}
		\label{fig:1-2-watrand}
	}
	\subfigure[Bio2RDF workloads]{
		\includegraphics[width=1.65in,height=1.48in]{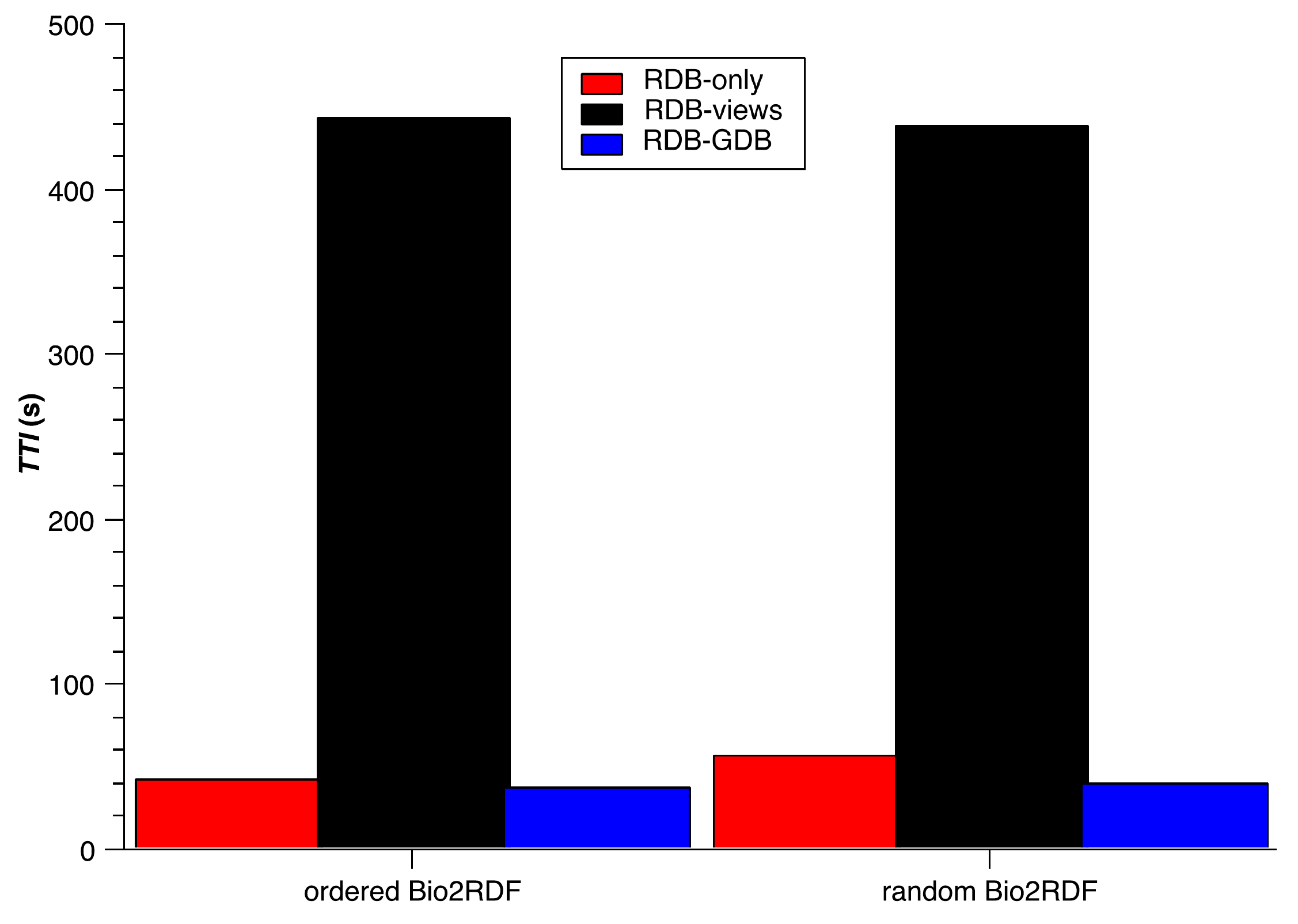}
		\label{fig:1-2-bio}
	}
\caption{Total Results of Each Workload Varying Store}
\label{fig:total-store}
\end{figure}

To explore the effectiveness of our proposed dual-store structure, we first compare our store \textsf{RDB-GDB} with two store variants, \textsf{RDB-only} and \textsf{RDB-views}. \textsf{RDB-only} stores and queries knowledge graphs in a relational database. To speed up the knowledge graph query process in a relational database, \textsf{RDB-views} creates the intermediate results of most frequent subqueries in the historical workloads as materialized views during the offline phase after processing a batch of queries. \textsf{RDB-GDB} leverages a graph store to accelerate the complex subquery process in a relational database and adopts \emph{DOTIL} to tune the physical design of dual-store structure periodically. For a fair comparison, the storage budgets for views in \textsf{RDB-views} and graph database in \textsf{RDB-GDB} are equal. To warm up the views in \textsf{RDB-views} and the graph database in \textsf{RDB-GDB}, we run each test 6 times and record the average \emph{TTI} of last 5 times. The comparison results of each batch on ordered workloads and random workloads are depicted in Figure~\ref{fig:batch-store-order} and Figure~\ref{fig:batch-store-rand}, respectively. The total results of each workload are shown in Figure~\ref{fig:total-store}.

As depicted in Figure~\ref{fig:batch-store-order} and Figure~\ref{fig:batch-store-rand}, \textsf{RDB-GDB} achieves better efficiency than \textsf{RDB-only} and \textsf{RDB-views} in all cases. The reason is that with the help of periodic physical design tuning, the triple partitions in the graph store of \textsf{RDB-GDB} effectively improve the complex subquery costs in \textsf{RDB-only}. Also, due to the storage constraints, the triple partitions selected by Q-learning model in \textsf{RDB-GDB} are more valuable than the views simply chosen based on frequency in \textsf{RDB-views}. Additionally, it can be observed that \emph{TTI} of \textsf{RDB-views} is sometimes higher than that of \textsf{RDB-only}. This is due to the expensive costs of looking up the appropriate views and joining view-table and other relational tables to generate final results. We also observe that \emph{TTI} of \textsf{RDB-GDB} is more stable than that of \textsf{RDB-only} and \textsf{RDB-views}. This is because that the dual-store tuner \emph{DOTIL} in \textsf{RDB-GDB} cumulates the historical experience of tuning valuable triple partitions. As the number of training knowledge graph queries increases, the performance of \emph{DOTIL} becomes better. From Figure~\ref{fig:total-store}, we observe that it makes little difference to the total \emph{TTI} of \textsf{RDB-GDB} on random and ordered workloads. The reason is that the adaptivity of \emph{DOTIL} to dynamic workloads is insusceptible to the query order. Compared with the most commonly used knowledge graph store \textsf{RDB-only}, \textsf{RDB-GDB} improves the query performance up to average 43.72\%. Compared with \textsf{RDB-views}, \textsf{RDB-GDB} improves the query performance up to average 63.01\%. This further demonstrates the limited effectiveness of relational views in accelerating complex query process.

Thus, we can conclude that our proposed dual-store structure achieves better performance in processing complex knowledge graph queries.

\subsection{Performance of \emph{DOTIL}}
\label{subsec:dotil}
To verify the performance of our proposed physical design tuner in the dual-store structure, we test \emph{DOTIL} along three dimensions: ($i$) tuning parameters; ($ii$) cold start of graph store; ($iii$) utilizing graph store with limited spare resources.

\subsubsection{Tuning Parameters}
In \emph{DOTIL}, there are five parameters, i.e. $r_{B_G}$ (the ratio of storage constraint $B_G$ to an entire knowledge graph), $prob$ (the initial probability of transferring), $\alpha$ (learning rate), $\gamma$ (discount factor), and $\lambda$ (the cost ratio in relational store to graph store). To determine the optimal values for these parameters, we measure \emph{TTI} and Q-matrix on half of the random version of YAGO workload varying each parameter. When we vary one parameter, the other parameters are set to their default values in Table~\ref{table:default}. Experimental results are shown in Table~\ref{table:paratuning}.

\begin{table}[!htb]
	\centering
	\caption{Default Values}
	\label{table:default}
	\begin{tabular}{cc}
		\hline
		\textbf{parameter} & \textbf{default value}   \\
		\hline
		$r_{B_G}$ &  25\%  \\
		$prob$ & 50\% \\
		$\alpha$ &  0.5   \\
		$\gamma$ &  0.5  \\
		$\lambda$ & 3.5 \\
		\hline
	\end{tabular}
\end{table}

As shown in Table~\ref{table:paratuning}, we vary $r_{B_G}$ from 20\% to 40\%. It can be observed that when $r_{B_G}$ is 25\%, \emph{TTI} is the least and the values in Q-matrix are the highest. When $r_{B_G}$ is less than 25\%, it is difficult for graph store to accommodate desired triple partitions and leads to poor training results. When $r_{B_G}$ is larger than 25\%, it is infrequent to evict the triple partitions whose Q values are low, hence the Q-matrix values decrease. Since the value of $prob$ is positively related to the number of Q-learning model training, we prefer to set $prob$ not less than 50\%. Therefore, the value of $prob$ is varied from 50\% to 100\%. We observe that \emph{TTI} fluctuates slightly because \emph{TTI} is unaffected by the value of $prob$. When $prob$ is 90\%, the values in Q-matrix are the highest. When $prob$ is less than 90\%, the number of model training decreases, and hence the Q-matrix values become less. We vary $\alpha$ from 0.3 to 0.7, and see that when $\alpha$ is 0.5, \emph{TTI} is the least and the values in Q-matrix are the highest. When $\alpha$ is less than 0.5, \emph{DOTIL} reduces the learning ability of new queries, and the training effects get worse. When $\alpha$ is more than 0.5, \emph{DOTIL} weakens its memory ability of historical experience and deteriorates the training results. The value of $\gamma$ is varied from 0.5 to 0.9, and we observe that when $\gamma$ is 0.7, \emph{TTI} is the least and the values in Q-matrix are the highest. When $\gamma$ is less than 0.7, \emph{DOTIL} overlooks the importance of previous experience and worsens the training results. When $\gamma$ is larger than 0.7, \emph{DOTIL} undervalues the instant reward which affects the Q values. Also, we vary $\lambda$ from 3.0 to 5.0, and see that when $\lambda$ is 4.5, \emph{TTI} is the least, and the values in Q-matrix are the highest. When $\lambda$ is less than 4.5, the difference between the cost in the graph store and that in relational store is diminished, and hence the reduced reward value affects the Q-matrix values. When $\lambda$ is more than 4.5, the training time of Q-learning model is too long to tune the physical design of dual-store structure efficiently.

Thus, to achieve high efficiency and well training performance of \emph{DOTIL}, we set 25\% as the value of $r_{B_G}$, 90\% as the value of $prob$, 0.5 as the value of $\alpha$, 0.7 as the value of $\gamma$, and 4.5 as the value of $\lambda$.

\begin{table}[!htb]
	\centering
	\caption{Results of Parameter Tuning}
	\label{table:paratuning}
	\begin{tabular}{c|ccc}
		\hline
		\textbf{parameter} & \textbf{value} & \textbf{\emph{TTI}} & \textbf{Q-matrix} \\
		\hline
		  & 20\% & 684.0760 & [0, 3.8263, 13.7780, 0] \\
		  & \textbf{25\%} & \textbf{653.4101} & \textbf{[0, 5.2717, 13.7956, 0]}  \\
		$r_{B_G}$ & 30\% & 674.3867 & [0, 3.8146, 9.8068, 0] \\
		  & 35\% & 700.0032 & [0, 6.5456, 6.3852, 0] \\
		  & 40\% & 674.8967 & [0, 3.4930, 11.7358, 0] \\
		\hline
		   & 50\% & 575.0997 & [0, 2.1057, 12.4025, 0] \\
		   & 60\% & 598.9008 & [0, 7.1411, 16.0713, 0] \\
		$prob$ & 70\% & 598.4994 & [0, 9.7398, 13.9126, 0] \\
		   & 80\% & 606.2928 & [0, 5.2241, 20.9603, 0] \\
		   & \textbf{90\%} & \textbf{607.7997} & \textbf{[0, 12.2230, 23.6434, 0]} \\
		   & 100\% & 609.6003 & [0, 7.0906, 18.3338, 0] \\
		  \hline
		   & 0.3 & 701.4354 & [0, 4.7951, 9.0349, 0] \\
		   & 0.4 & 663.7754 & [0, 4.2571, 16.0011, 0] \\
		$\alpha$ &  \textbf{0.5} & \textbf{654.8847} & \textbf{[0, 5.2816, 15.1309, 0]} \\
		  & 0.6 & 748.5279 & [0, 1.9856, 8.3675, 0] \\
		  & 0.7 & 702.1908 & [0, 0.0826, 1.4354, 0] \\
		 \hline
		  & 0.5 & 645.9246 & [0, 6.5384, 10.3453, 0] \\
		  & 0.6 & 695.1923 & [0, 1.6490, 12.5269, 0] \\
		$\gamma$ &  \textbf{0.7} & \textbf{598.7205} & \textbf{[0, 7.4015, 12.9472, 0]} \\
		 & 0.8 & 612.1578 & [0, 9.3671, 9.7828, 0]  \\
		 & 0.9 & 685.9546 & [0, 7.5959, 12.3609, 0]  \\
		 \hline
		 & 3.0 & 697.7290 & [0, 3.2530, 12.5419, 0] \\
		 & 3.5 & 604.4655 & [0, 4.7346, 18.8977, 0] \\
		$\lambda$ & 4.0 & 608.4625 & [0, 5.0882, 22.0954, 0] \\
		 & \textbf{4.5} & \textbf{608.4121} & \textbf{[0, 5.6235, 25.1015, 0]} \\
		 & 5.0 & 605.9493 & [0, 6.5287, 22.0688, 0] \\
		\hline
	\end{tabular}
\end{table}

\subsubsection{Cold Start of Graph Store}
\begin{figure}[htb!]
	\centering
	\subfigure[ordered YAGO workloads]{
		\includegraphics[width=1.65in,height=1.48in]{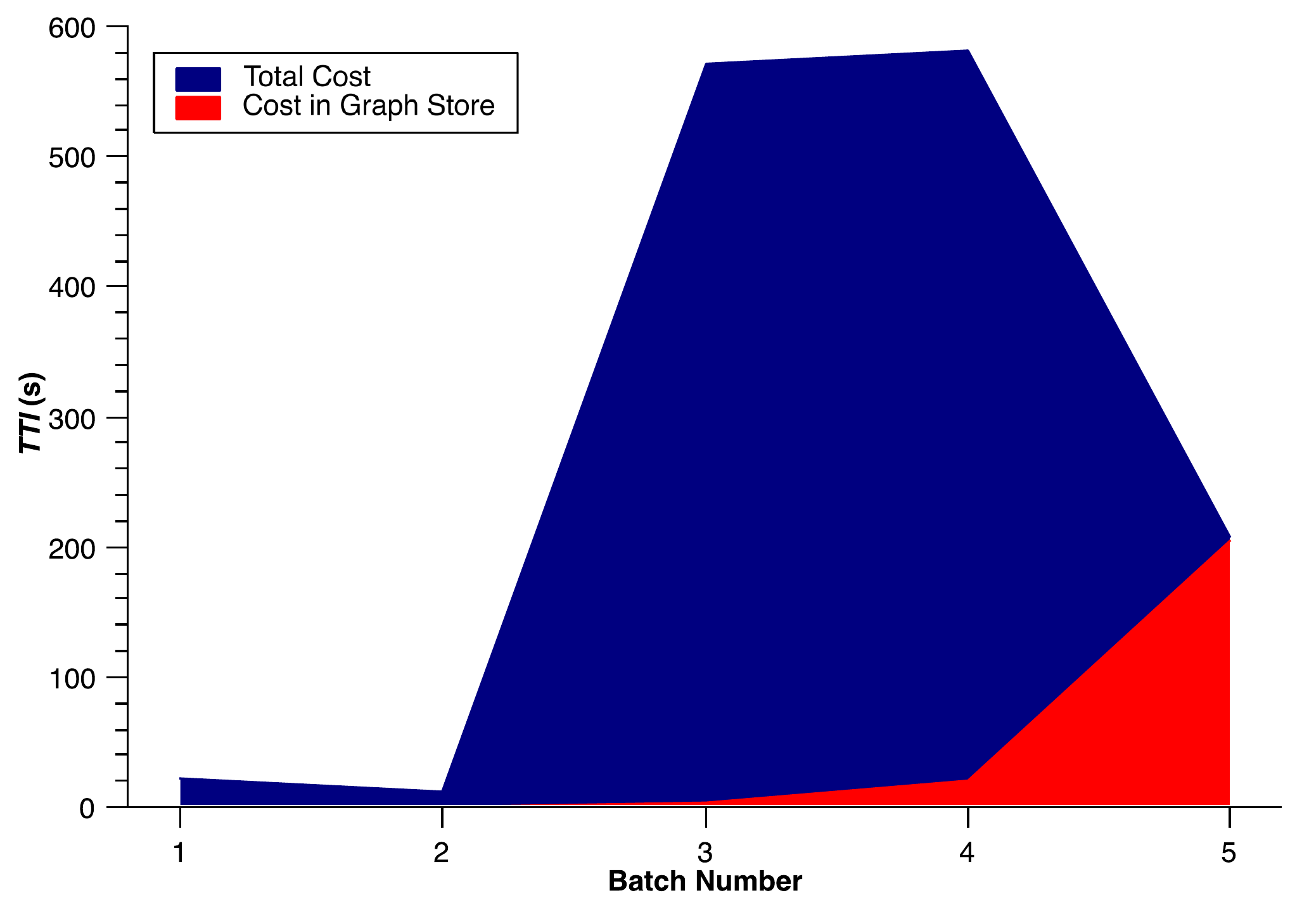}
		\label{fig:2-1-order}
	}
	\subfigure[random YAGO workloads]{
		\includegraphics[width=1.65in,height=1.48in]{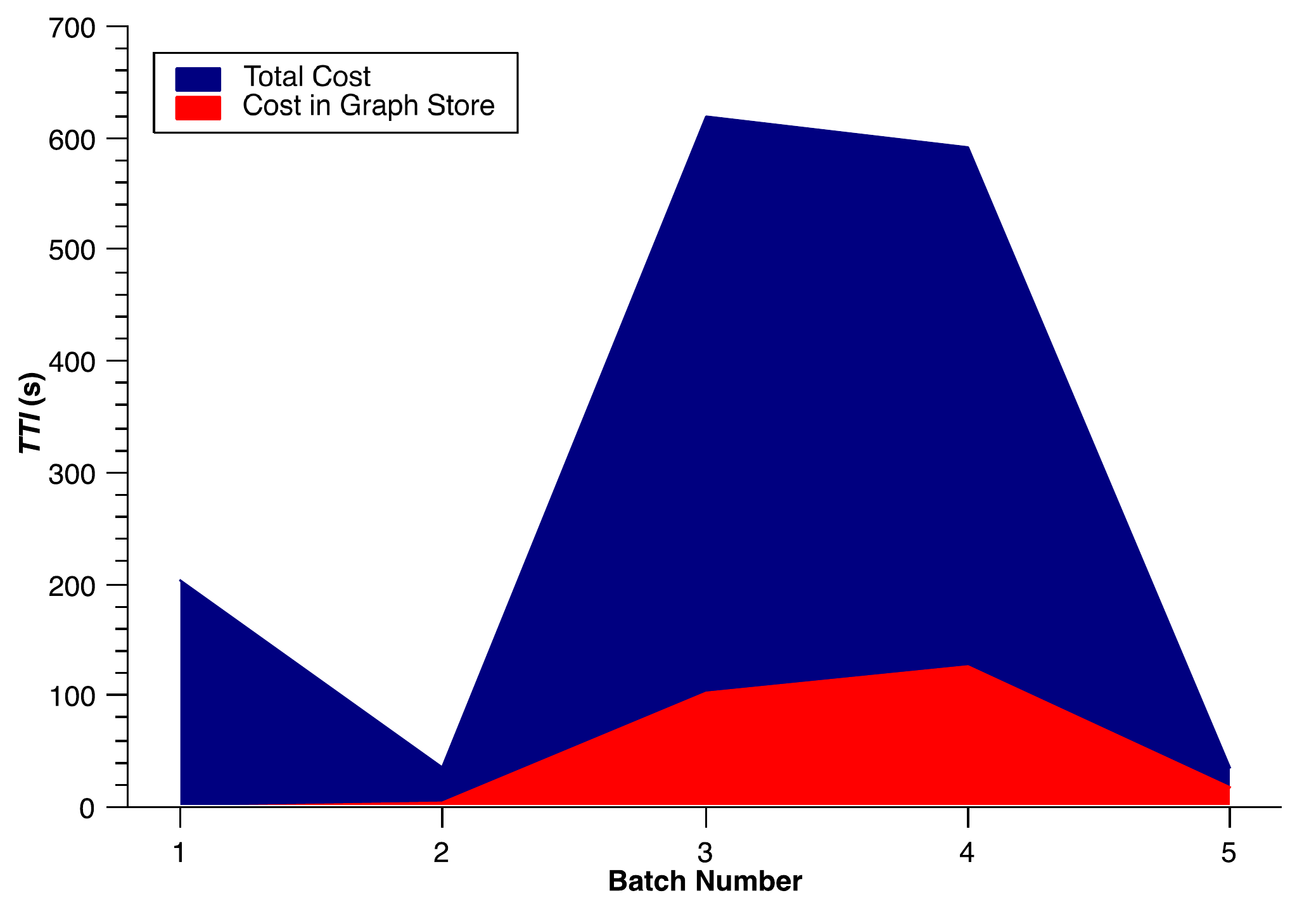}
		\label{fig:2-1-random}
	}
	\caption{Cost Proportion of Graph Store}
	\label{fig:coldstart}
\end{figure}

In the initial state of dual-store structure, the graph store is empty. As \emph{DOTIL} tunes the dual-store physical design, the triple partitions in the graph store gradually increase. Thus, there is a cold start of graph store, and its influence on the overall performance needs to be tested. To test this influence, we measure \emph{TTI} of both the cost in the graph store and the total cost from the initial state of the graph store. Experimental results are depicted in Figure~\ref{fig:coldstart}.

From Figure~\ref{fig:coldstart}, we observe that although in the first two batches, the cost proportion of the graph store is small, it rises rapidly from the third batch. This is due to the fact that \emph{DOTIL} transfers valuable triple partitions to the graph store in a short time, and hence the increased new queries are processed in the graph store. Thus, we conclude that the cold start of graph store has little impact on the performance of our dual-store structure.

\subsubsection{Utilizing Graph Store with Limited Space Resources}
\begin{figure}[htb!]
	\centering
	\subfigure[IO consumed of graph store]{
		\includegraphics[width=1.65in,height=1.48in]{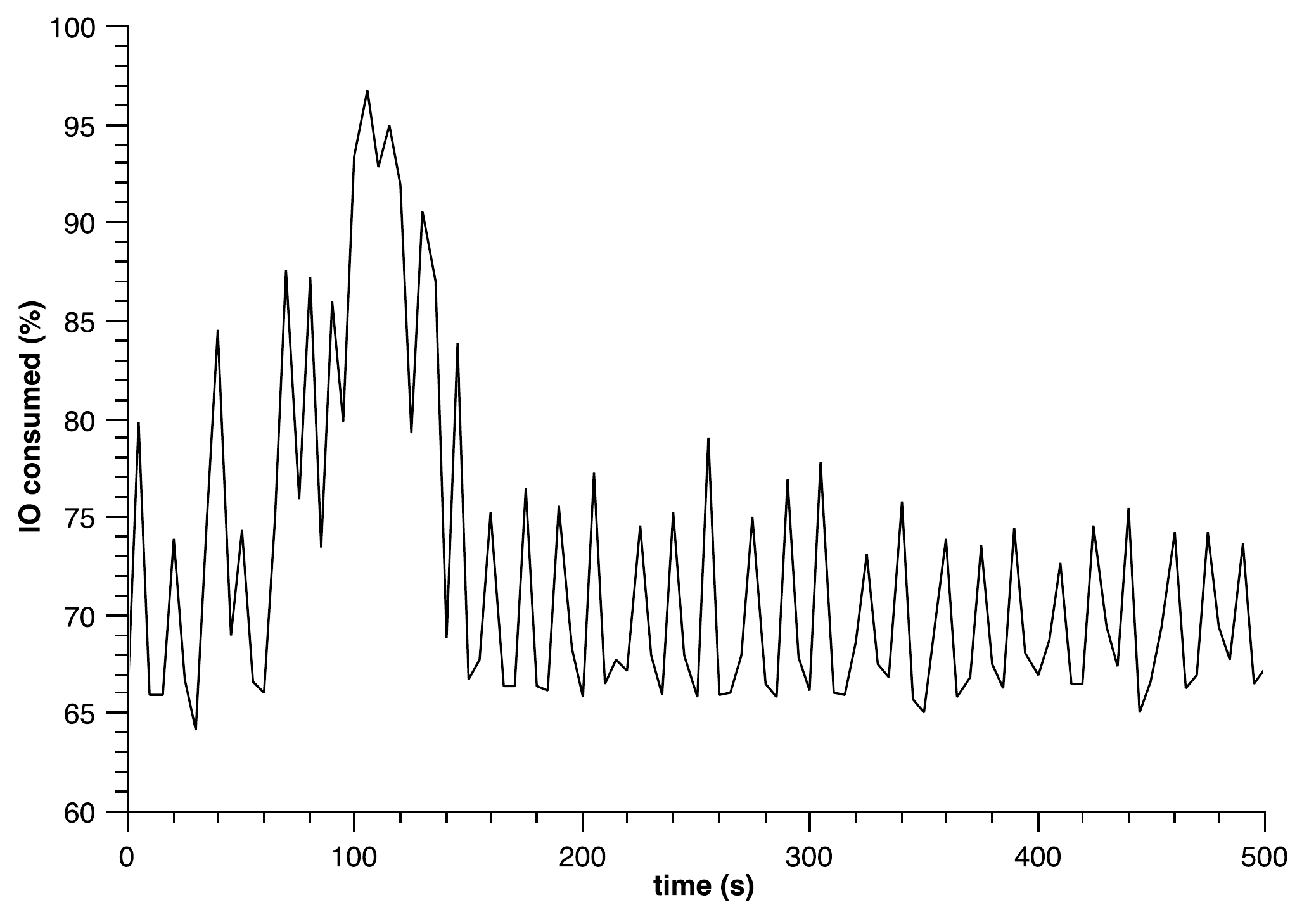}
		\label{fig:2-2-io}
	}
	\subfigure[CPU consumed of graph store]{
		\includegraphics[width=1.65in,height=1.48in]{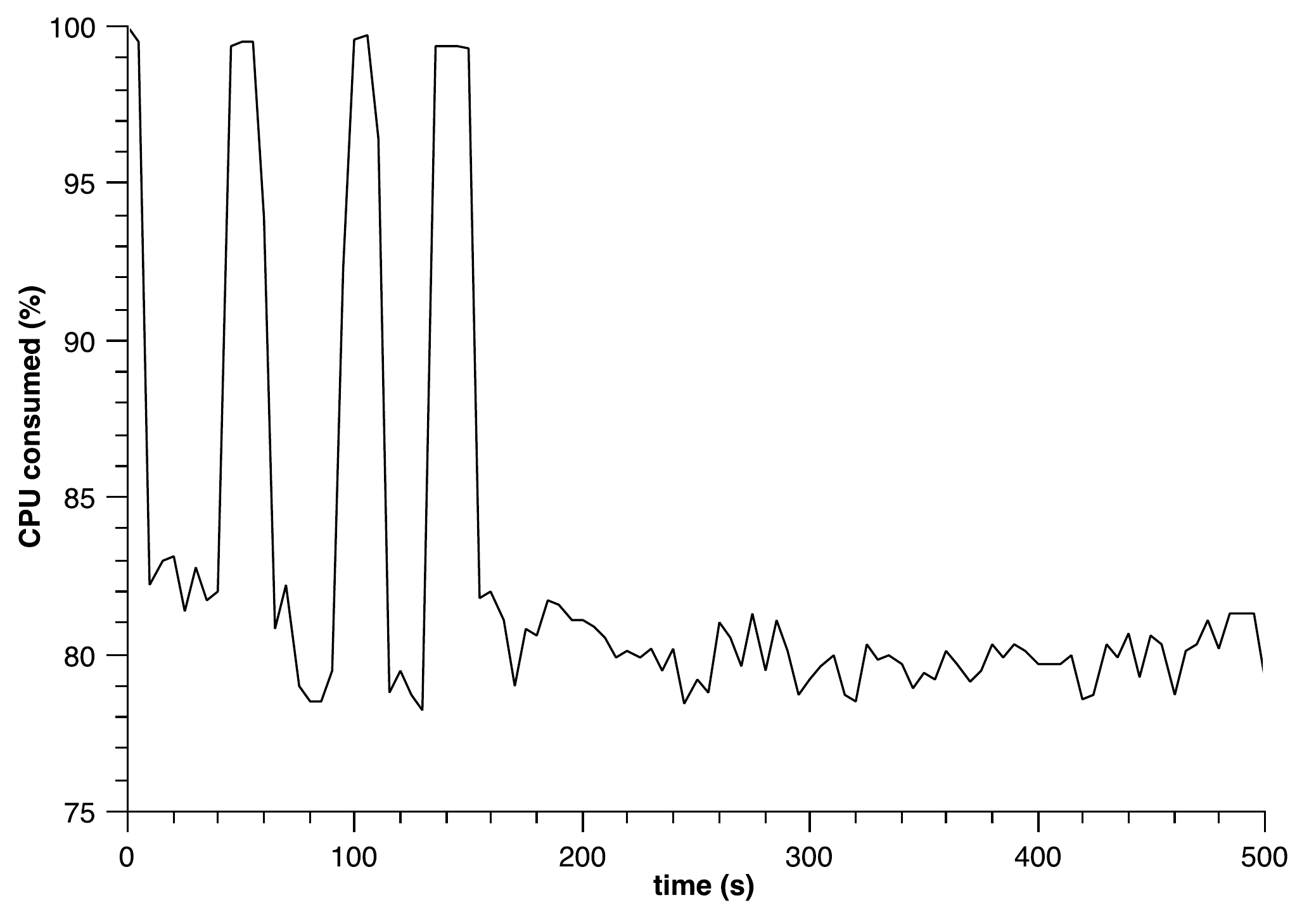}
		\label{fig:2-2-cpu}
	}
	\caption{Impacts of Parallel Thread on Resources Consumed of Graph Store}
	\label{fig:consumed}
\end{figure}

In the process of \emph{DOTIL}, there is a parallel thread to compute the query execution time in the relational store. In the meantime, we process queries in the graph store. Since the resources are partly used by the parallel thread, we study the impacts of limited spare resources on utilizing graph store. We first test the slowdown of graph store with 40\% IO, 20\% IO, 40\% CPU, and 20\% CPU, and report the results in Table~\ref{table:slowdown}. Then, with 40\% IO resource, we measure the time-varying percent of IO and CPU consumed in the graph store. Experimental results are depicted in Figure~\ref{fig:consumed}.

As shown in Table~\ref{table:slowdown}, it can be observed that the slowdown percent of the graph store is little in spite of few spare resources. From Figure~\ref{fig:consumed}, we can also see that the percent of consumed IO or CPU fluctuates widely in the beginning, and it then stabilizes at a small value. Therefore, the influence of limited resources on graph store utilization is little.

\begin{table}[!htb]
	\centering
	\caption{Slowdown of Graph Store with Limited Spare Resources}
	\label{table:slowdown}
	\begin{tabular}{cc}
		\hline
		\textbf{spare resource} & \textbf{slowdown percent}   \\
		\hline
		IO 40\% &  0.45\% \\
		IO 20\% & 0.30\% \\
		CPU 40\% &  5\%   \\
		CPU 20\% &  18\%  \\
		\hline
	\end{tabular}
\end{table}

\subsection{Comparisons of Tuners}
\label{subsec:tuners}
\begin{figure}[htb!]
	\centering
	\subfigure[YAGO workloads]{
		\includegraphics[width=1.65in,height=1.48in]{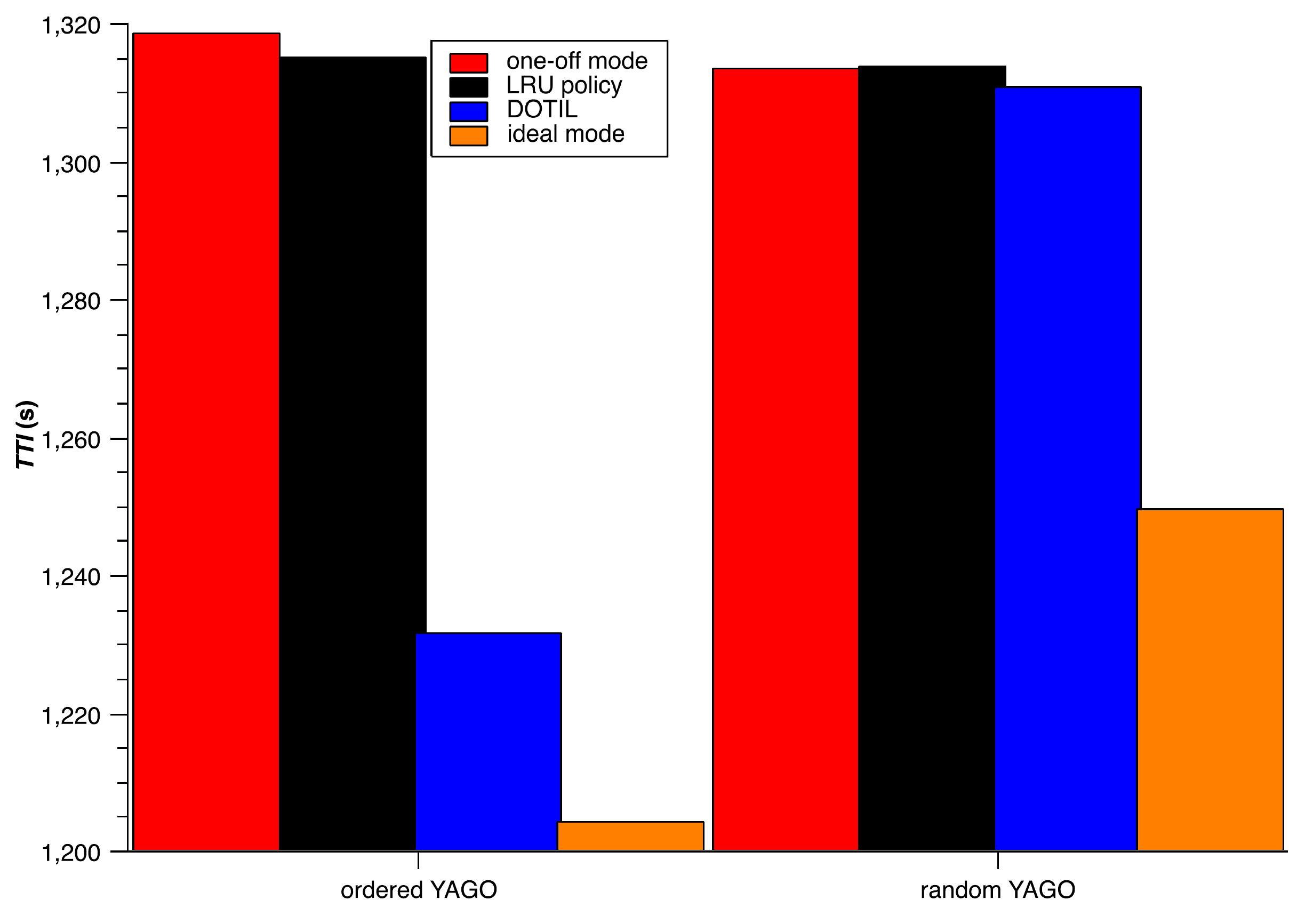}
		\label{fig:3-1-yago}
	}
	\subfigure[ordered WatDiv workloads]{
		\includegraphics[width=1.65in,height=1.48in]{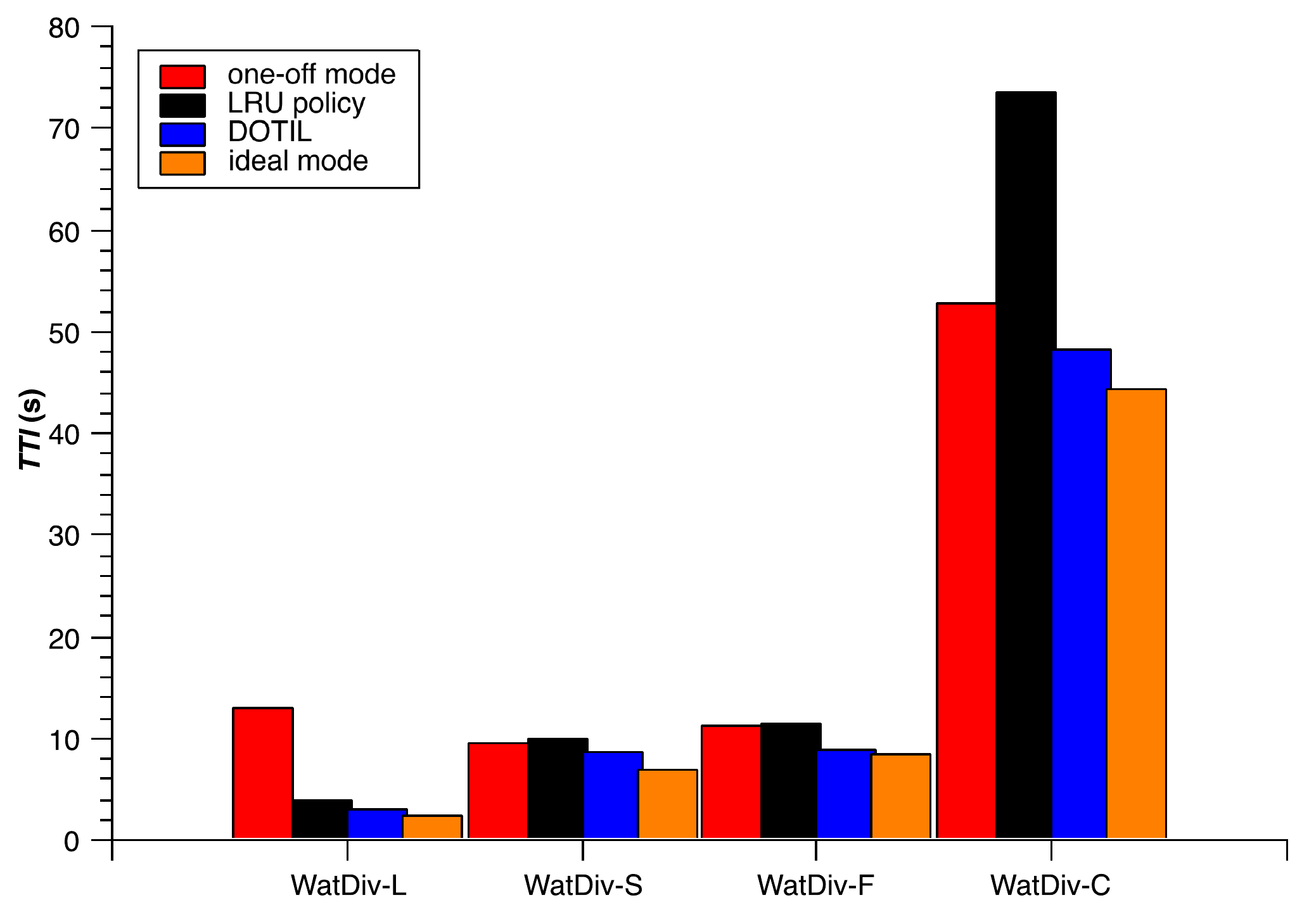}
		\label{fig:3-1-watorder}
	}
	\subfigure[random WatDiv workloads]{
		\includegraphics[width=1.65in,height=1.48in]{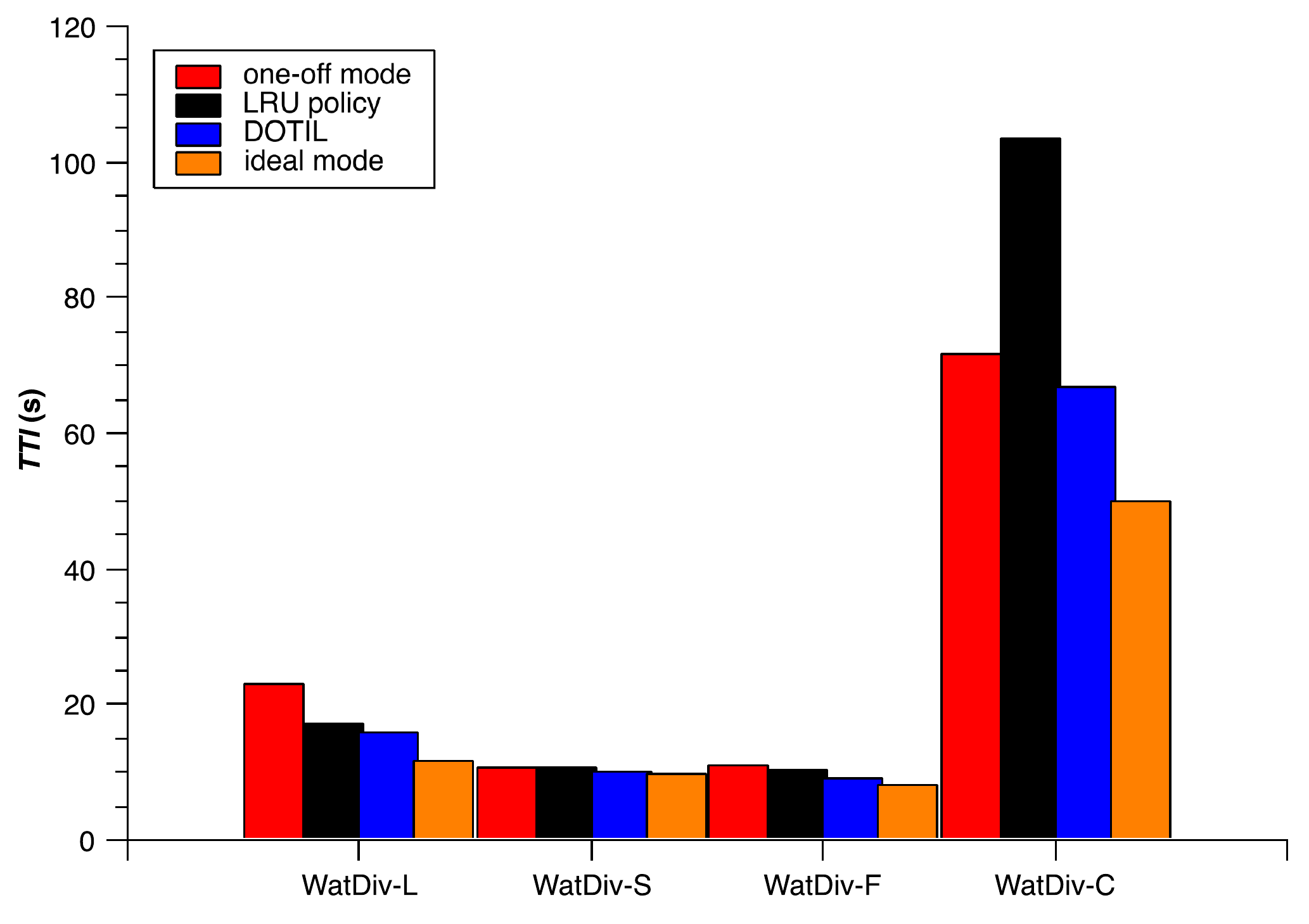}
		\label{fig:3-1-watrand}
	}
	\subfigure[Bio2RDF workloads]{
		\includegraphics[width=1.65in,height=1.48in]{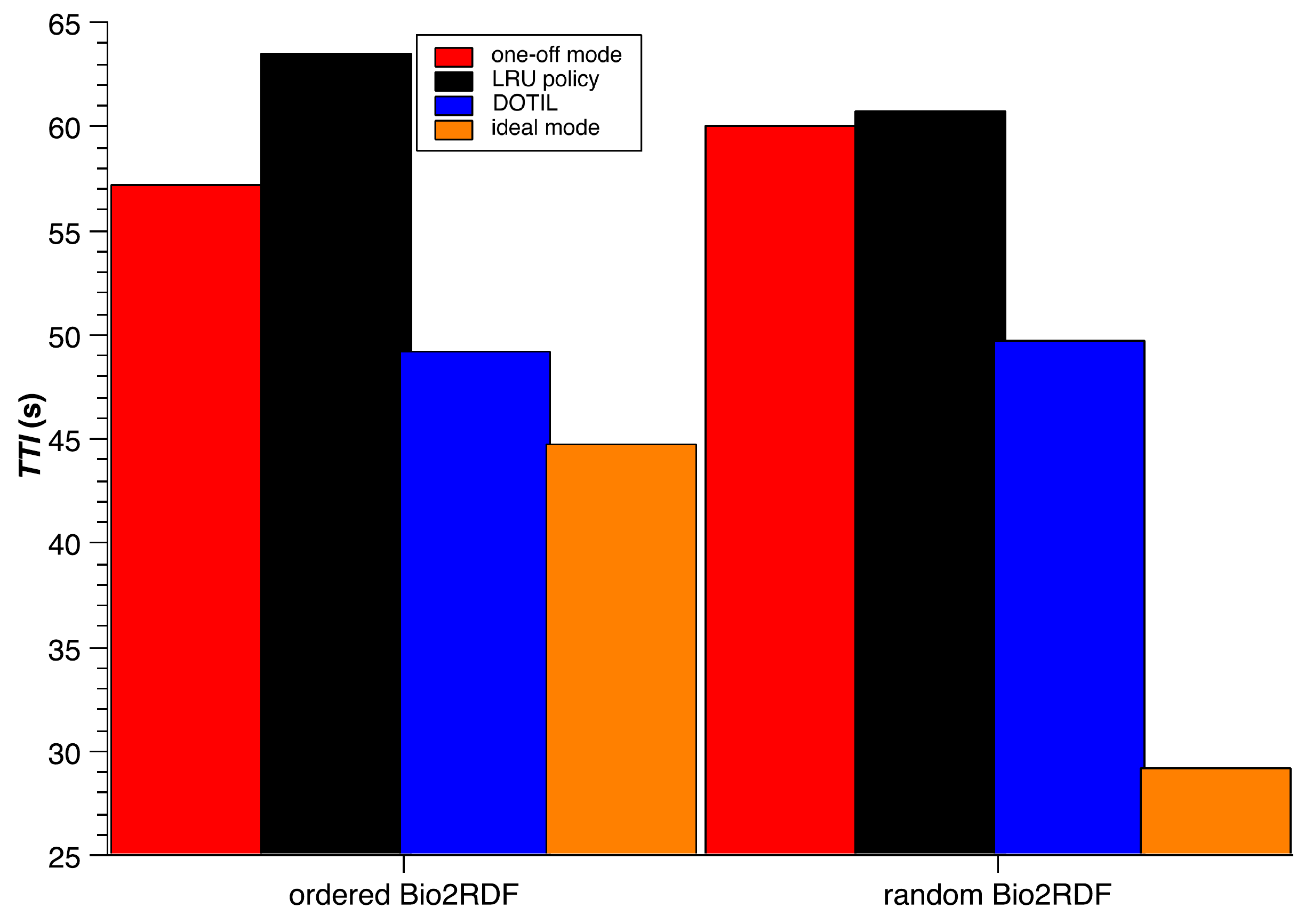}
		\label{fig:3-1-bio}
	}
	\caption{Comparison Results with Other Tuners}
	\label{fig:compare-tuners}
\end{figure}

To evaluate the effectiveness of our proposed physical design tuner \textsf{DOTIL}, we compare \textsf{DOTIL} with other three tuning methods, \textsf{one-off mode}, \textsf{LRU policy}, and \textsf{ideal mode}. \textsf{One-off mode} foresees the whole future workload and tunes the dual-store structure once at the beginning time. \textsf{LRU policy} transfers the most frequent triple partitions in the historical workloads to the graph store after each batch. \textsf{Ideal mode} foresees the workload in next batch and tunes the dual-store structure beforehand. Note that \textsf{ideal mode} is the ideal situation of \textsf{DOTIL}, we evaluate the difference between \textsf{DOTIL} and its ideal case. To warm up the graph database, we test \emph{TTI} 6 times and report the average of last 5 times. The comparison results are illustrated in Figure~\ref{fig:compare-tuners}.

From Figure~\ref{fig:compare-tuners}, we observe that \emph{TTI} of \textsf{DOTIL} is significantly less than that of \textsf{one-off mode} and \textsf{LRU policy}. The reason is that \textsf{one-off mode} keeps the limited valuable triple partitions in the graph store due to the storage constraint. Its \emph{static} property makes it lack the adaptivity to dynamic workloads. The most frequent triple partitions chosen by \textsf{LRU policy} may not bring most benefits, and hence leads to high time costs. Additionally, it can be seen that on ordered workloads, the difference between \textsf{DOTIL} and \textsf{ideal mode} is less than that on random workloads. This is because that mutations of a query template are clustered in ordered workloads, and \emph{DOTIL} is more adaptive to similar queries than irrelevant ones.

The evaluation results confirm the high effectiveness of \emph{DOTIL}. With the dynamic physical design tuning of \emph{DOTIL}, the dual-store structure achieves satisfactory performances.

\section{Conclusion}
\label{sec:conclude}
In this paper, we propose a novel dual-store structure for knowledge graphs. We leverage a graph store to accelerate complex knowledge graph query process in traditional relational stores. To determine the triple partitions which are valuable to transfer from relational store to graph store, we derive a reinforcement-learning-based tuner \textsf{DOTIL}. With \textsf{DOTIL}, the dual-store structure adapts to the dynamic changing workloads. Our experimental results on real knowledge graphs verify the effectiveness and efficiency of our proposed dual-store structure. In the future, we plan to extend the storage structure so that it can be applied to various challenging scenarios.



\ifCLASSOPTIONcompsoc
  \section*{Acknowledgments}
\else
  \section*{Acknowledgment}
\fi

This paper was partially supported by NSFC grant  U1866602, 61602129, 61772157,  CCF-Huawei Database System Innovation Research Plan DBIR2019005B and Microsoft Research Asia.



\bibliographystyle{IEEEtran}
\bibliography{main.bbl}

\begin{thebibliography}{10}
\providecommand{\url}[1]{#1}
\csname url@samestyle\endcsname
\providecommand{\newblock}{\relax}
\providecommand{\bibinfo}[2]{#2}
\providecommand{\BIBentrySTDinterwordspacing}{\spaceskip=0pt\relax}
\providecommand{\BIBentryALTinterwordstretchfactor}{4}
\providecommand{\BIBentryALTinterwordspacing}{\spaceskip=\fontdimen2\font plus
\BIBentryALTinterwordstretchfactor\fontdimen3\font minus
  \fontdimen4\font\relax}
\providecommand{\BIBforeignlanguage}[2]{{%
\expandafter\ifx\csname l@#1\endcsname\relax
\typeout{** WARNING: IEEEtran.bst: No hyphenation pattern has been}%
\typeout{** loaded for the language `#1'. Using the pattern for}%
\typeout{** the default language instead.}%
\else
\language=\csname l@#1\endcsname
\fi
#2}}
\providecommand{\BIBdecl}{\relax}
\BIBdecl

\bibitem{dbpedia}
``Dbpedia,'' \url{https://wiki.dbpedia.org/about}.

\bibitem{linkedgeodata}
``Linkedgeodata,'' \url{http://www.linkedgeodata.org/About}.

\bibitem{uniprot}
``Uniprot,'' \url{https://www.uniprot.org/help/about}.

\bibitem{neo4j}
``Neo4j,'' \url{https://neo4j.com/docs/developer-manual/current/}.

\bibitem{zou2014gstore}
L.~Zou, M.~T. {\"O}zsu, L.~Chen, X.~Shen, R.~Huang, and D.~Zhao, ``gstore: a
  graph-based sparql query engine,'' \emph{The VLDB journal}, vol.~23, no.~4,
  pp. 565--590, 2014.

\bibitem{Rodriguez2012The}
M.~A. Rodriguez and P.~Neubauer, ``The graph traversal pattern,'' \emph{Graph
  Data Management Techniques \& Applications}, 2012.

\bibitem{3store}
S.~Harris and G.~Nicholas, ``3store: Efficient bulk rdf storage,'' in
  \emph{Proceedings of the 1st International Workshop on Practical and Scalable
  Semantic Systems}, 2004, pp. 81--95.

\bibitem{DLDB}
Z.~Pan and J.~Heflin, ``Dldb: Extending relational databases to support
  semantic web queries,'' in \emph{Proceedings of the 1st International
  Workshop on Practical and Scalable Semantic Systems}, 2004, pp. 109--113.

\bibitem{Jena}
K.~Wilkinson, ``Jena property table implementation,'' in \emph{Proceedings of
  the 2nd International Workshop on Scalable Semantic Web Knowledge Base
  Systems}, 2006, pp. 35--46.

\bibitem{abadi2009sw}
D.~J. Abadi, A.~Marcus, S.~R. Madden, and K.~Hollenbach, ``Sw-store: a
  vertically partitioned dbms for semantic web data management,'' \emph{The
  VLDB Journal}, vol.~18, no.~2, pp. 385--406, 2009.

\bibitem{neumann2008rdf}
T.~Neumann and G.~Weikum, ``Rdf-3x: a risc-style engine for rdf,''
  \emph{Proceedings of the VLDB Endowment}, vol.~1, no.~1, pp. 647--659, 2008.

\bibitem{weiss2008hexastore}
C.~Weiss, P.~Karras, and A.~Bernstein, ``Hexastore: sextuple indexing for
  semantic web data management,'' \emph{Proceedings of the VLDB Endowment},
  vol.~1, no.~1, pp. 1008--1019, 2008.

\bibitem{sun2015sqlgraph}
W.~Sun, A.~Fokoue, K.~Srinivas, A.~Kementsietsidis, G.~Hu, and G.~Xie,
  ``Sqlgraph: An efficient relational-based property graph store,'' in
  \emph{Proceedings of the 2015 ACM SIGMOD International Conference on
  Management of Data}, 2015, pp. 1887--1901.

\bibitem{pokorny2017integrity}
J.~Pokorný, M.~Valenta, and J.~Kovacic, ``Integrity constraints in graph
  databases,'' \emph{Procedia Computer Science}, vol. 109, pp. 975--981, 2017.

\bibitem{bruno2007online}
N.~Bruno and S.~Chaudhuri, ``An online approach to physical design tuning,'' in
  \emph{2007 IEEE 23rd International Conference on Data Engineering}.\hskip 1em
  plus 0.5em minus 0.4em\relax IEEE, 2007, pp. 826--835.

\bibitem{schnaitter2007line}
K.~Schnaitter, S.~Abiteboul, T.~Milo, and N.~Polyzotis, ``On-line index
  selection for shifting workloads,'' in \emph{2007 IEEE 23rd International
  Conference on Data Engineering Workshop}.\hskip 1em plus 0.5em minus
  0.4em\relax IEEE, 2007, pp. 459--468.

\bibitem{schnaitter2012semi}
K.~Schnaitter and N.~Polyzotis, ``Semi-automatic index tuning: Keeping dbas in
  the loop,'' \emph{Proceedings of the VLDB Endowment}, vol.~5, no.~5, 2012.

\bibitem{consens2012divergent}
M.~P. Consens, K.~Ioannidou, J.~LeFevre, and N.~Polyzotis, ``Divergent physical
  design tuning for replicated databases,'' in \emph{Proceedings of the 2012
  ACM SIGMOD International Conference on Management of Data}, 2012, pp. 49--60.

\bibitem{lefevre2014miso}
J.~LeFevre, J.~Sankaranarayanan, H.~Hacigumus, J.~Tatemura, N.~Polyzotis, and
  M.~J. Carey, ``Miso: souping up big data query processing with a multistore
  system,'' in \emph{Proceedings of the 2014 ACM SIGMOD international
  conference on Management of data}, 2014, pp. 1591--1602.

\bibitem{bugiotti2015invisible}
F.~Bugiotti, D.~Bursztyn, A.~Deutsch, I.~Ileana, and I.~Manolescu, ``Invisible
  glue: Scalable self-tuning multi-stores,'' in \emph{Conference on Innovative
  Data Systems Research (CIDR)}, 2015.

\bibitem{horowitz1974computing}
E.~Horowitz and S.~Sahni, ``Computing partitions with applications to the
  knapsack problem,'' \emph{Journal of the ACM}, vol.~21, no.~2, pp. 277--292,
  1974.

\bibitem{van2017automatic}
D.~Van~Aken, A.~Pavlo, G.~J. Gordon, and B.~Zhang, ``Automatic database
  management system tuning through large-scale machine learning,'' in
  \emph{Proceedings of the 2017 ACM International Conference on Management of
  Data}, 2017, pp. 1009--1024.

\bibitem{kara2018columnml}
K.~Kara, K.~Eguro, C.~Zhang, and G.~Alonso, ``Columnml: Column-store machine
  learning with on-the-fly data transformation,'' \emph{Proceedings of the VLDB
  Endowment}, vol.~12, no.~4, pp. 348--361, 2018.

\bibitem{kraska2018case}
T.~Kraska, A.~Beutel, E.~H. Chi, J.~Dean, and N.~Polyzotis, ``The case for
  learned index structures,'' in \emph{Proceedings of the 2018 International
  Conference on Management of Data}, 2018, pp. 489--504.

\bibitem{ding2019ai}
B.~Ding, S.~Das, R.~Marcus, W.~Wu, S.~Chaudhuri, and V.~R. Narasayya, ``Ai
  meets ai: Leveraging query executions to improve index recommendations,'' in
  \emph{Proceedings of the 2019 International Conference on Management of
  Data}, 2019, pp. 1241--1258.

\bibitem{ghrab2016grad}
A.~Ghrab, O.~Romero, S.~Skhiri, A.~A. Vaisman, and E.~Zimanyi, ``Grad: On graph
  database modeling,'' \emph{arXiv: Databases}, 2016.

\bibitem{yago}
``Yago,''
  \url{https://www.mpi-inf.mpg.de/departments/databases-and-information-systems/research/yago-naga/yago/}.

\bibitem{watdiv}
``Watdiv,'' \url{https://dsg.uwaterloo.ca/watdiv/}.

\bibitem{biordf}
``Bio2rdf,'' \url{https://download.bio2rdf.org/}.

\bibitem{harbi2016accelerating}
R.~Harbi, I.~Abdelaziz, P.~Kalnis, N.~Mamoulis, Y.~Ebrahim, and M.~Sahli,
  ``Accelerating sparql queries by exploiting hash-based locality and adaptive
  partitioning,'' \emph{The VLDB Journal}, vol.~25, no.~3, pp. 355--380, 2016.

\bibitem{wattemp}
``Watdiv query templates,''
  \url{https://dsg.uwaterloo.ca/watdiv/basic-testing.shtml}.

\bibitem{Michael2011Big}
M.~S. Hopkins, ``Big data, analytics and the path from insights to value,''
  \emph{MIT Sloan Management Review}, vol.~52, no.~2, pp. 21--22, 2011.

\end{thebibliography}
%

%

\end{document}